\documentclass[fleqn,usenatbib]{rasti}

\usepackage{newtxtext,newtxmath}

\usepackage[T1]{fontenc}
\usepackage{graphicx}%
\usepackage{subcaption}
\usepackage{caption}
\graphicspath{{figs_revised/}}
\usepackage{bm,csquotes}
\usepackage{booktabs}
\usepackage{float}
\usepackage{tabularx}
\usepackage{siunitx}
\usepackage{makecell}
\usepackage{array}
\usepackage{geometry} 
\usepackage{xcolor}   
\usepackage{colortbl}

\DeclareRobustCommand{\VAN}[3]{#2}
\let\VANthebibliography\thebibliography
\def\thebibliography{\DeclareRobustCommand{\VAN}[3]{##3}\VANthebibliography}

\usepackage{graphicx}	
\usepackage{amsmath}	
\usepackage{changepage}

\title[FIR Filter Design for GRAO Beam Optimisation]{Digital Beam Pattern Optimisation for the GRAO 32-m Telescope: A Comparative Analysis of FIR Filter Design Methods}

\author[ T. Ansah-Narh et al.]{
Theophilus  Ansah-Narh,$^{1}$\thanks{E-mail: \href{theophilus.ansah-narh@gaec.gov.gh}{theophilus.ansah-narh@gaec.gov.gh} (TA-N)}
Nia Imara,$^{2}$
Benedicta Woode$^{1}$
and Emmanuel Proven Adzri$^{1}$
\\
$^{1}$Ghana Space Science and Technology Institute, Ghana Atomic Energy Commission, P.O. Box LG 80, Legon-Accra, Ghana\\
$^{2}$Astronomy and Astrophysics Department, University of California, Santa Cruz, CA, USA\\
}

\date{Accepted XXX. Received YYY; in original form ZZZ}

\pubyear{\the\year{}}

\begin{document}
\label{firstpage}
\pagerange{\pageref{firstpage}--\pageref{lastpage}}
\maketitle

\begin{abstract}
The scientific utility of large single-dish radio telescopes depends critically on the stability and fidelity of their beam patterns, which govern angular resolution, sensitivity, and polarimetric accuracy. For the 32‑m Ghana Radio Astronomy Observatory (GRAO) antenna, electromagnetic simulations reveal residual sidelobes, structural diffraction, and cross‑polar leakage that limit performance in high‑dynamic‑range and polarisation‑sensitive observations. To address these limitations, we develop a finite‑impulse‑response (FIR) spatial filtering framework that reformulates beam optimisation as a digital signal processing problem. By exploiting the equivalence between angular displacement and spatial frequency, classical FIR design methods window‑based and Parks–McClellan algorithms are adapted to operate directly on simulated Jones fields. This approach enables controlled suppression of high spatial frequency artefacts responsible for sidelobes and polarisation mixing, while preserving the telescope’s diffraction‑limited resolution. Applied to the GRAO $5$ GHz beam model, the method achieves substantial reductions in near‑in sidelobe ripple, improves beam smoothness, and lowers cross‑polar leakage below $-30$ dB at boresight. These improvements translate into enhanced calibration stability and polarimetric precision, strengthening the telescope’s capacity for Very Long Baseline Interferometry, spectral‑line surveys, and pulsar timing. Beyond GRAO, the method provides a generalisable, non‑invasive, and computationally efficient pathway for beam control applicable to other single‑dish and phased‑array instruments. The results establish digital spatial filtering as a practical complement to conventional optical or mechanical optimisation, advancing the integration of electromagnetic modelling and signal processing in next‑generation radio astronomical instrumentation.
\end{abstract}




\section{Introduction} \label{sec:intro}

The Ghana Radio Astronomy Observatory (GRAO) represents a significant advancement in Africa’s growing capabilities in radio astronomy and a key milestone in the development of the African Very Long Baseline Interferometry (AVN) network. 
The AVN aims to establish a network of single-dish radio telescopes across the continent, thereby enriching the coverage and sensitivity of global VLBI arrays when integrated with the European VLBI Network and the wider international community \citep{Atemkeng2022,gaylard2014african}.
In converting the former Kutunse telecommunications antenna into a dedicated radio telescope, GRAO provides a foundational node in this pan-African astronomical infrastructure. 
Its equatorial location (approx.\ 5.75$^{\circ}$ N, 0.31$^{\circ}$ W) affords both favourable sky coverage of the Milky Way’s galactic plane and a strategically important position for VLBI baselines that include Africa \citep{Atemkeng2022}.

Beyond its strategic geographical value, GRAO supports capacity-building objectives. The project has stimulated training of engineers and scientists within Ghana and the region, aligning with broader initiatives such as the Development in Africa with Radio Astronomy (DARA\footnote{\url{https://www.dara-project.org/}}) programme, which fosters human capital development across Sub-Saharan Africa in preparation for major facilities such as the Square Kilometre Array (SKA) \citep{Hoare2018}.
As a functioning radio telescope, GRAO enables a diverse suite of scientific programmes including methanol maser monitoring, pulsar timing and broader survey science, positioning the facility as a valuable contributor both locally and collaboratively \citep{2026JATIS..12a7001P, GRAO2023}.

To fully realise this scientific potential, however, optimisation of the telescope’s fundamental observational characteristics is required. The radiation (far-field) beam pattern of any radio telescope governs its angular resolution, sidelobe behaviour and overall sensitivity. In practice, real systems deviate from the ideal aperture illumination due to surface errors, structural blockage, feed illumination non-idealities and mechanical tolerances: these lead to sidelobe structures, beam asymmetries and degraded polarisation performance. For single-dish instruments in particular, such imperfections introduce systematic errors in flux measurements, polarimetric analyses and time-domain studies of compact sources.

Uncertainties in beam shape propagate directly into errors in derived astrophysical parameters, especially when observing extended emission or in crowded fields \citep{thompson2017interferometry}. In polarimetric observations, beam asymmetries and non-ideal cross-polar responses may generate spurious polarisation or distort measured polarisation angles \citep{Warnick2012, wijnholds2012polarimetry}.
For GRAO’s scientific goals, including accurate maser flux variability studies or precision pulsar timing, a well-characterised and optimised beam is therefore essential to minimise systematic bias and maximise observational reliability.

Traditional mitigation approaches typically involve mechanical adjustment of the feed or reflector, illumination tapering, physical blocking of support-structure diffraction, or tailored observing strategies (e.g.\ off-axis calibration). While effective to some extent, such techniques often lack flexibility once the telescope is operational, and are not easily adapted for different observational requirements or changing system components.

Digital signal-processing (DSP) techniques provide a powerful and complementary path. 
In particular, finite-impulse‐response (FIR) filters are extensively used in radio-astronomy back-ends for spectral channelisation, RFI mitigation and time-domain signal conditioning, owing to their stability, linear-phase behaviour and controlled frequency responses \citep{price2021spectrometers,pfister2017discrete}.
Extending this concept, one can exploit a spatial–frequency duality: by mapping angular displacement from boresight into a normalised spatial frequency variable under small-angle approximation, the angular (or radial) beam-pattern of a dish can be treated analogously to a temporal/frequency filter design problem.
Under this mapping, FIR filter coefficients correspond to effective spatial weighting of the aperture illumination (or far-field angular samples) and classical filter-design methods (e.g.\ window-based methods, Parks–McClellan equiripple designs) can be repurposed for beam-shape optimisation.
While spatial filtering techniques are well established in array and phased-array systems \citep{Wang2025, Garakoui2023, liu2010wideband}, their application in optimising single-dish beam patterns (via post-capture weighting of measured or simulated beam grids) remains relatively unexplored.

In this paper, we present a comprehensive analysis of FIR-filter applications for beam-pattern optimisation of the GRAO 32-m telescope. We utilise high-resolution electromagnetic simulation data (via the \texttt{GRASP}\footnote{\url{https://www.ticra.com/software/grasp/}} package) sampled on a spherical grid (elevation: $201$ points, 0$^\circ$--3$^\circ$; azimuth: $181$ points, 0$^\circ$--360$^\circ$) with full Jones-matrix entries ($J_{qh}, J_{qv}, J_{ph}, J_{pv}$) to assess spatial weighting strategies. We evaluate filter designs using astronomy-relevant metrics including half-power beamwidth (HPBW), peak sidelobe level (dB), beam efficiency and polarimetric cross-polarisation leakage, and we compare multiple FIR design techniques (e.g.\ window families, equiripple optimisation). Furthermore, we discuss implementation trade-offs including filter length (spatial taps), computational cost and real-world integration in the GRAO signal chain. Our contributions are: 

\begin{adjustwidth}{1.5em}{0pt}
\begin{enumerate}[i.]
    \item the formal mapping of classical FIR filter-design methods into the angular domain of a single-dish telescope, 
    \item a quantitative comparison of multiple FIR spatial-filter designs applied to GRAO beam data, and 
    \item practical recommendations for the deployment of such filters in astronomical back-ends.
\end{enumerate} 
\end{adjustwidth}

\noindent
We believe that optimising GRAO’s beam pattern in this way will enhance its scientific output and strengthen its role within the AVN and the global astronomical community.

The paper is structured as follows: Section~\ref{sec:grao_telescope} describes the GRAO telescope system, the simulation framework and the beam-pattern data structure; Section~\ref{sec:fir} details the FIR-design mapping and optimisation methodology; Section~\ref{sec:beam_opt_results} presents results of the filter designs and their impacts on the beam metrics; Section~\ref{sec:discussion_implications} discusses implementation issues, limitations and future extensions; finally, Section~\ref{sec:conclusions} summarises our findings and offers concluding remarks.

\section{The GRAO 32-m Telescope System}
\label{sec:grao_telescope}

\subsection{Technical Specifications and System Architecture}
\label{subsec:specs_design}

The technical capabilities of the GRAO 32-m telescope that underpin this beam pattern analysis are summarised in Table~\ref{tab:grao_spec}. These specifications represent the baseline performance metrics against which the beam optimisation techniques are evaluated, with particular focus on the 5~GHz operational band that forms the basis of this study.

\begin{table*}
\centering
\caption{Technical Specifications of the GRAO 32\,m Telescope}
\label{tab:grao_spec}
\resizebox{\textwidth}{!}{%
\begin{tabular}{|l|l|}
\hline
\textbf{Category} & \textbf{Specification} \\
\hline
\multicolumn{2}{|c|}{\textbf{Location}} \\
\hline
Latitude/Longitude & \SI{5.750485}{\degree N}, \SI{-0.305116}{\degree W} \\
\hline
\multicolumn{2}{|c|}{\textbf{Antenna Structure}} \\
\hline
Mount Type & Alt-azimuth (wheel-and-track) \\
Optics & Shaped Cassegrain with beam-waveguide (2 concave, 2 flat mirrors) \\
Diameter & \SI{32}{\meter} \\
Height (zenith/horizon) & \SI{38.3}{\meter} / \SI{37.5}{\meter} \\
Movable Mass & \SI{230}{\tonne} \\
Azimuth Range & $\pm$\SI{300}{\degree} (relative to North) \\
Elevation Range & \SI{7}{\degree} to \SI{90}{\degree} \\
Surface Accuracy (RMS) & \SI{1.88}{\milli\meter} \\
Main Reflector Panels & 240 (thickness: \SI{1.6}{\milli\meter}) \\
\hline
\multicolumn{2}{|c|}{\textbf{Performance Parameters}} \\
\hline
Max Gain & \SI{63.72}{\dB} (5 GHz), \SI{66.40}{\dB} (6.7 GHz) \\
Aperture Efficiency & 81\% (5 GHz), 77\% (6.7 GHz)\\
Half-Power Beamwidth (HPBW) & \SI{0.11}{\degree} (5 GHz), \SI{0.09}{\degree} (6.7 GHz)\\
Sidelobe Level & \SI{-15.21}{\dB} (5 GHz), \SI{-15.15}{\dB} (6.7 GHz)\\
Spillover Loss & \SI{0.19}{\dB} (5 GHz), \SI{0.14}{\dB} (6.7 GHz)\\
Cross-Pol Isolation & \SI{-31.88}{\dB} (5 GHz), \SI{-32.21}{\dB} (6.7 GHz)\\
System Temperature ($T_{\text{sys}}$) & \SI{125}{\kelvin} (5 GHz), \SI{<90}{\kelvin} (6.7 GHz)\\
Sensitivity ($A_{\text{eff}}/T_{\text{sys}}$) & 
\SI{5.21}{\square\meter\per\kelvin} (5 GHz), 
\SI{4.95}{\square\meter\per\kelvin} (6.7 GHz)\\
\hline
\multicolumn{2}{|c|}{\textbf{Receiver System}} \\
\hline
Polarisation & Dual circular (L + R) \\
Cooling & Ambient (uncooled) \\
Noise Temperature & \SI{110}{\kelvin} (5 GHz), \SI{<90}{\kelvin} (6.7 GHz) \\
Local Oscillator Tuning & \SI{1}{\hertz} resolution \\
Frequency Range & \SI{4.917}{}--\SI{5.045}{\GHz} (5 GHz band), \SI{6.55}{}--\SI{6.95}{\GHz} (6.7 GHz band) \\
\hline
\multicolumn{2}{|c|}{\textbf{Dynamic Performance}} \\
\hline
Slew Rate (Azimuth/Elevation) & \SI{0.27}{\degree\per\second} / \SI{0.29}{\degree\per\second} \\
Acceleration (Azimuth/Elevation) & \SI{0.056}{\degree\per\square\second} / \SI{0.072}{\degree\per\square\second} \\
Tracking Accuracy & \SI{0.05}{\degree} RMS \\
\hline
\multicolumn{2}{|c|}{\textbf{Environmental Tolerance}} \\
\hline
Wind Speed Operating Limit & \SI{15}{\meter\per\second} (accuracy: \SI{\pm 0.3}{\meter\per\second}) \\
Temperature Range & \SI{-40}{\degreeCelsius} to \SI{+60}{\degreeCelsius} (accuracy: \SI{\pm 0.5}{\degreeCelsius}) \\
\hline
\end{tabular}%
}
\end{table*}

The antenna employs a shaped Cassegrain configuration with beam-waveguide optics, comprising two concave and two flat mirrors that guide signals to the receiver cabin. This design achieves a surface accuracy of \SI{1.88}{\milli\meter} RMS, which is particularly notable given the telescope's conversion from telecommunications use. The shaped optics optimise aperture efficiency while controlling spillover and sidelobe levels, achieving efficiencies exceeding 77\% across both primary operating bands \citep{2026JATIS..12a7001P, GRAO2023}. The wheel-and-track alt-azimuth mount provides $\pm$\SI{300}{\degree} azimuth range and \SI{7}{\degree} to \SI{90}{\degree} elevation coverage, with tracking accuracy of \SI{0.05}{\degree} RMS, sufficient for most single-dish observing programmes and VLBI requirements.

The receiver system employs dual circular polarisation with uncooled receivers achieving system temperatures below \SI{125}{\kelvin} at 5~GHz and below \SI{90}{\kelvin} at 6.7~GHz. The measured cross-polarisation isolation better than \SI{-31}{\dB} across both bands indicates excellent polarisation purity, which is crucial for polarimetric studies and VLBI observations \citep{thompson2017interferometry}. The combination of high aperture efficiency and low system temperature yields sensitivity metrics ($A_{\text{eff}}/T_{\text{sys}}$) of \SI{5.21}{\square\meter\per\kelvin} and \SI{4.95}{\square\meter\per\kelvin} at 5~GHz and 6.7~GHz respectively, positioning the telescope competitively for its size class.

\subsection{GRASP simulation framework} \label{sec:grasp}

\begin{figure}
	\centering
	\includegraphics[width=\columnwidth]{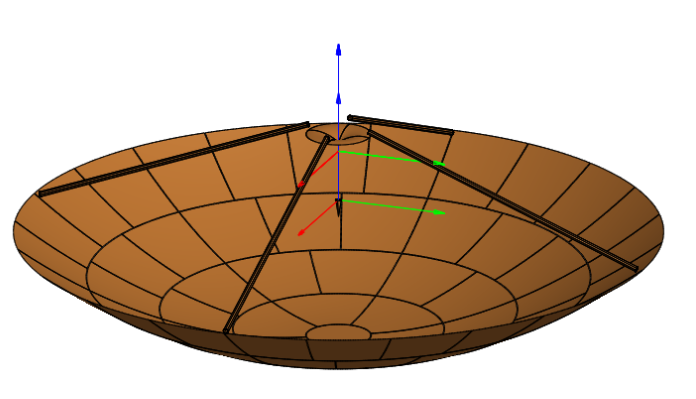}
    \caption{Geometrical model of a parabolic reflector antenna showing the segmented surface mesh and the primary feed support structure. The coordinate systems (red, green, and blue axes) indicate the global and local phase centres for the optical analysis.}
    \label{fig:Fig_1}
\end{figure}

Electromagnetic simulations of the telescope beam were conducted using the \texttt{GRASP10.6} software suite, employing the physical optics (PO) and physical theory of diffraction (PTD) modules.
Figure~\ref{fig:Fig_1} shows the geometrical model of the GRAO 32‑m antenna used in these simulations, illustrating the segmented surface mesh and the primary feed support structure.
This approach is particularly well-suited to electrically large dual-reflector systems such as the 32-m dish described in Section~\ref{subsec:specs_design}.
In the PO approximation, the induced surface current on the reflector is expressed as
\begin{equation}
\mathbf{J}(\mathbf{r}) = 2\,\hat{\mathbf{n}}\times \mathbf{H}_{\rm inc}(\mathbf{r}), 
\label{eq:surface_current}
\end{equation}
where \(\hat{\mathbf{n}}\) denotes the local surface normal and \(\mathbf{H}_{\rm inc}\) is the incident magnetic field from the feed. The far-field electric field is then obtained by integrating these currents over the reflector surface according to
\begin{equation}
\mathbf{E}(\theta,\phi) = \frac{j k}{2\pi}\,e^{-jkr}\frac{e^{-jkr}}{r}\iint_{A_{\rm refl}} \mathbf{J}(\mathbf{r}')\,e^{\,j k \hat{\mathbf{r}}\!\cdot\!\mathbf{r}'}\,\mathrm{d}A'.
\label{eq:farfield_integral}
\end{equation}
Here \(k=2\pi/\lambda\) is the free-space wavenumber, \(\hat{\mathbf{r}}\) the observation direction, and \(A_{\rm refl}\) the illuminated reflector surface.

To support full polarisation characterisation, the simulation was set up with two orthogonal feed excitations (h, v). For each excitation the far-field components \(E_{\theta}\) and \(E_{\phi}\) were computed for both incident polarisation states. These results are assembled into the complex Jones matrix at each angular coordinate \((\theta,\phi)\),
\begin{equation}
\mathbf{J}(\theta,\phi) =
\begin{bmatrix}
J_{qh}(\theta,\phi) & J_{qv}(\theta,\phi)\\
J_{ph}(\theta,\phi) & J_{pv}(\theta,\phi)
\end{bmatrix}
\equiv
\begin{bmatrix}
E_{\theta,\mathrm{h}} & E_{\theta,\mathrm{v}}\\
E_{\phi,\mathrm{h}}     & E_{\phi,\mathrm{v}}
\end{bmatrix},
\label{eq:jones_matrix}
\end{equation}
where the subscripts ‘$h$’ and ‘$v$’ denote horizontal and vertical feed excitations and the first index of \(J\) corresponds to the output field component. This formalism enables a complete description of the polarimetric response of the system, including amplitude and phase relationships between polarisation channels \citep{hamaker1996understanding}.

The simulated beam-pattern grid comprises four complex matrix elements ($J_{qh}, J_{qv}, J_{ph}, J_{pv}$) sampled on a spherical coordinate grid: polar angle \(\theta\) spanning \(0^\circ\) to \(3^\circ\) in increments of \(0.015^\circ\) ($201$ points) and azimuth angle \(\phi\) spanning \(0^\circ\) to \(360^\circ\) in steps of \(2^\circ\) ($181$ points). 
The grid resolution was selected to ensure adequate sampling of the main beam (approximately $7$ points across HPBW at $5$ GHz) and to identify asymmetries in the sidelobe region. The data structure and indexing convention follow those commonly used in radio-astronomy instrumentation literature \citep{Smirnov2011}.

These simulated Jones-matrix beam maps form the foundational input for the spatial weighting and filter-design analysis described in Section \ref{sec:fir}. In particular, the grid data serve as the \enquote{time-domain} equivalent in our FIR spatial-filter framework, enabling the adaptation of classical filter-design methods (window functions, Parks–McClellan equiripple algorithms) to beam-pattern optimisation.

\section{FIR-Based Beam Pattern Optimisation Framework}  \label{sec:fir}
\subsection{Mathematical Framework} \label{sec:fir_math}

Building upon the electromagnetic formulation presented in Section~\ref{sec:grasp}, the far-field response of the telescope is represented by the simulated Jones matrix $\mathbf{J}(\theta,\phi)$, whose elements describe the complex polarimetric gain pattern of the antenna.  
Each Jones component can be viewed as a discretised spatial sequence sampled uniformly in elevation angle $\theta$, encoding both amplitude and phase variations across the beam.  
The objective here is to reformulate the beam-optimisation problem within a DSP framework by constructing a FIR spatial filter that operates directly on these sampled fields to suppress high–spatial–frequency artefacts (sidelobes) while minimally perturbing the main lobe.

Let $\{E_i[n]\}$ denote the discrete angular samples of the complex field corresponding to the $i^{\mathrm{th}}$ Jones component, where the index $n$ refers to uniform increments $\theta_n = n\,\Delta\theta$.  
A linear, shift-invariant spatial filtering operation is expressed as
\begin{equation}
\widetilde{E}_i[n] = 
\sum_{k=0}^{M-1} h[k]\,E_i[n-k],
\label{eq:spatial_conv}
\end{equation}
where $h[k]$ are the real-valued filter coefficients of order $M$.  
Equation~\eqref{eq:spatial_conv} performs angular convolution analogous to temporal filtering in DSP, with the independent variable here representing spatial displacement rather than time.  

The spectral behaviour of the filter is characterised by its discrete Fourier transform
\begin{equation}
H(\omega_\theta) = \sum_{k=0}^{M-1} h[k]\,e^{-j\omega_\theta k},
\label{eq:spatial_response}
\end{equation}
where $\omega_\theta$ is the normalised angular frequency proportional to the physical wavenumber,
\begin{equation}
\omega_\theta = \frac{2\pi}{\lambda}\sin\theta \equiv k_\theta .
\label{eq:freq_wavenumber}
\end{equation}
This relation establishes the frequency–wavenumber equivalence central to the present approach: manipulating the angular distribution of power in $k_\theta$ space is formally identical to shaping the frequency spectrum of a discrete-time signal \citep{alkin2025discrete,price2021spectrometers,oppenheim1999discrete}.  
Accordingly, the design of $h[k]$ governs how spatial frequencies contribute to the reconstructed beam, providing a mathematically rigorous route to controlled sidelobe attenuation and main-beam refinement.

The idealised spatial response for beam smoothing can be expressed as a low-pass function,
\begin{equation}
|H_{\mathrm{ideal}}(\omega_\theta)| =
\begin{cases}
1, & |\omega_\theta| \le \omega_c,\\[4pt]
0, & \omega_c < |\omega_\theta| \le \pi,
\end{cases}
\label{eq:ideal_lp}
\end{equation}
where the cut-off spatial frequency $\omega_c$ is selected to correspond approximately to the telescope's half-power beamwidth (HPBW).  
High-$k_\theta$ components, responsible for small-scale oscillations and far-out sidelobes, are therefore suppressed, while the coherent main-beam structure remains largely intact.

A linear-phase constraint $h[k] = h[M-1-k]$ is imposed to ensure constant group delay across all spatial frequencies.  
This guarantees that the relative phase between the polarisation components of $\mathbf{J}(\theta,\phi)$ is preserved, preventing distortion of the polarimetric calibration terms.  
The resulting filtered field $\widetilde{E}_i(\theta)$ thus retains the physical coherence of the original simulation while exhibiting a smoother, better-confined angular response.

The subsequent subsections develop two complementary strategies for realising the FIR coefficients: a window-based construction and an optimal equiripple (Parks–McClellan) design, followed by the quantitative performance metrics adopted for evaluating beam improvements.

\subsection{Window-Based FIR Construction} \label{sec:window_fir}

The impulse response corresponding to the ideal low-pass filter in Eq.~\eqref{eq:ideal_lp} is the discrete sinc function
\begin{equation}
h_{\mathrm{ideal}}[n] = 
\frac{\sin(\omega_c(n - (M-1)/2))}{\pi (n - (M-1)/2)}.
\label{eq:sinc_function}
\end{equation}
Because this infinite sequence must be truncated to finite length $M$, spectral leakage and Gibbs oscillations appear in the angular response.  
To mitigate these effects, a window function $w[n]$ is applied:
\begin{equation}
h[n] = h_{\mathrm{ideal}}[n]\,w[n].
\label{eq:windowed}
\end{equation}
The window $w[n]$ acts as an apodisation operator that smooths discontinuities at the filter edges, analogous to tapering in aperture synthesis \citep{harris2005use}.  

Several classical window families were evaluated--Hamming, Hann, Blackman, Bartlett, rectangular, and Kaiser--each defining a distinct compromise between main-lobe width $\Delta\theta_\mathrm{mb}$ and sidelobe attenuation $A_\mathrm{sl}$.  
For instance, the Hamming window yields $A_\mathrm{sl}\approx -43~\mathrm{dB}$ and $\Delta\theta_\mathrm{mb}\approx 1.36\Delta\theta_\mathrm{ideal}$, while the Blackman window suppresses sidelobes below $-58~\mathrm{dB}$ but with $\Delta\theta_\mathrm{mb}$ broadened by $\sim 1.7$.
Particularly flexible is the Kaiser window, defined as
\begin{equation}
w[n] = \frac{I_0\!\left(\beta \sqrt{1-\left(\frac{2n}{M-1}-1\right)^2}\right)}{I_0(\beta)},
\quad 0 \le n \le M-1,
\label{eq:kaiser}
\end{equation}
where $I_0(\cdot)$ is the zeroth-order modified Bessel function.  
The shape parameter $\beta$ controls the trade-off between stopband attenuation and transition width: larger $\beta$ enhances sidelobe suppression at the cost of main-beam broadening.  
For this study, $\omega_c$ was selected to correspond to the HPBW of the unfiltered GRAO beam ($\sim0.11^\circ$ at 5~GHz), and $\beta$ was tuned to minimise the integrated sidelobe power while maintaining $<10\%$ increase in HPBW.  
Filter coefficients were implemented using \textsc{SciPy}'s \texttt{firwin} routine with $M=65$ taps.

\subsection{Optimal Parks–McClellan Synthesis} \label{sec:pm_fir}

While windowed-sinc methods provide intuitive control, they do not guarantee optimality.  
The Parks–McClellan algorithm \citep{mcclellan2005personal,parks1972chebyshev} employs the Remez exchange technique to obtain the FIR filter that minimises the maximum deviation between the designed and desired responses under a Chebyshev norm.  
The objective function is defined as
\begin{equation}
\min_{h[k]} E_\infty = 
\max_{\omega_\theta \in \Omega}
W(\omega_\theta)
\left|H_{\mathrm{d}}(\omega_\theta) - 
\sum_{k=0}^{M-1} h[k]\,e^{-j\omega_\theta k}\right|,
\label{eq:remez}
\end{equation}
where $H_{\mathrm{d}}(\omega_\theta)$ is the ideal low-pass profile and $W(\omega_\theta)$ is a weighting function emphasising critical regions of the spectrum.  
Alternating error extrema of equal magnitude are enforced iteratively, leading to equiripple behaviour across both passband and stopband.  

In the beam-filtering context, the weighting function was set such that $W_{\mathrm{stop}}/W_{\mathrm{pass}} \approx 10$, enforcing stringent sidelobe suppression.  
Filter lengths of $M=10$, $20$, $40$, and $65$ taps were synthesised to examine how kernel length influences angular confinement and numerical stability.  
The optimised coefficients $h[k]$ were then convolved with each Jones component of Eq.~\eqref{eq:jones_matrix}, producing filtered beams $\widetilde{J}_{ab}(\theta,\phi)$ that exhibit reduced angular sidebands yet preserve the polarimetric phase relationships.

\subsection{Energy and Performance Metrics} \label{sec:metrics}

The impact of spatial filtering on the radiometric and geometric performance of the GRAO telescope was quantified through a set of standardised beam metrics derived from the far-field power distribution
\begin{equation}
P(\theta,\phi) = |E(\theta,\phi)|^2,
\label{eq:power_field}
\end{equation}
where $E(\theta,\phi)$ is the complex voltage field obtained either from the unfiltered Jones components or their filtered counterparts.  
Each metric provides insight into a specific physical attribute of the beam angular resolution, off-axis energy rejection, and power confinement, and together they form a consistent framework for assessing improvements achieved by FIR-based beam optimisation.

\paragraph*{\textnormal{Half-Power Beamwidth (HPBW):}}
The HPBW quantifies the telescope’s angular resolving capability.  
It is defined as the angular separation between the two points at which the normalised power response falls by 3~dB relative to the peak intensity:
\begin{equation}
\mathrm{HPBW} = \theta_2 - \theta_1, \qquad 
20\log_{10}\!\left(\frac{|E(\theta_i)|}{|E(0)|}\right)=-3.
\label{eq:hpbw}
\end{equation}
For an unperturbed circular aperture of diameter $D$, the theoretical HPBW approximates to $1.02\,\lambda/D$.  
Spatial filtering modifies the effective aperture illumination function, thereby slightly broadening the beam according to the passband width of $H(\omega_\theta)$.  
A well-designed filter should preserve $\mathrm{HPBW}$ within $\sim10$~per~cent of its nominal value while mitigating sidelobe oscillations.  
Excessive broadening, although beneficial for sidelobe suppression, results in reduced angular resolution and should be avoided in high-precision astrometric or VLBI applications.

\paragraph*{\textnormal{Peak Sidelobe Level (SLL):}}
The SLL measures the relative strength of the most prominent sidelobe with respect to the main-beam maximum:
\begin{equation}
\mathrm{SLL} = 
20\log_{10}\!\left(\frac{\max_{\theta>\theta_\mathrm{mb}} |E(\theta)|}{|E(0)|}\right),
\label{eq:sll}
\end{equation}
where $\theta_\mathrm{mb}$ defines the outer boundary of the main lobe, typically taken as the first null in $P(\theta)$.  
Sidelobes arise from diffraction and discontinuities in aperture illumination, leading to unwanted pickup of radiation from bright sources or the ground.  
In single-dish radio astronomy, such contamination directly impacts flux calibration accuracy and spectral line dynamic range.  
A key objective of FIR-based filtering is therefore to minimise $\mathrm{SLL}$ while maintaining high on-axis gain.  
Reduction of $\mathrm{SLL}$ by even a few decibels can substantially lower systematic contamination in extended-source mapping and continuum imaging.

\paragraph*{\textnormal{Main-Beam Efficiency ($\eta_\mathrm{mb}$):}}
The main-beam efficiency characterises the fraction of total radiated or received power that is contained within the principal lobe of the antenna pattern:
\begin{equation}
\eta_\mathrm{mb} =
\frac{\displaystyle \int_{\Omega_\mathrm{mb}} |E(\theta,\phi)|^2 \,\mathrm{d}\Omega}
{\displaystyle \int_{4\pi} |E(\theta,\phi)|^2 \,\mathrm{d}\Omega}.
\label{eq:eta_mb}
\end{equation}
Here, $\Omega_\mathrm{mb}$ denotes the solid angle of the main beam.  
For ideal reflectors with uniform illumination and negligible blockage, $\eta_\mathrm{mb}$ approaches unity, while in practical systems, scattering from structural supports and surface errors distributes part of the energy into sidelobes.  
Filtering effectively redistributes this energy back toward the main lobe, improving $\eta_\mathrm{mb}$ and hence the telescope’s aperture efficiency.  
This metric also serves as a proxy for radiometric sensitivity, since the system gain $G \propto \eta_\mathrm{mb} D^2/\lambda^2$ under fixed receiver conditions.

\paragraph*{\textnormal{Sidelobe Suppression Ratio (SSR):}}
A complementary global measure of beam confinement is the sidelobe suppression ratio,
\begin{equation}
\mathrm{SSR} = 
10\log_{10}
\!\left(
\frac{\displaystyle \int_{\Omega_\mathrm{mb}} |E|^2\,\mathrm{d}\Omega}
{\displaystyle \int_{\Omega_\mathrm{sl}} |E|^2\,\mathrm{d}\Omega}
\right),
\label{eq:ssr}
\end{equation}
where $\Omega_\mathrm{sl}$ represents the angular region encompassing all sidelobes.  
Unlike $\mathrm{SLL}$, which reflects the peak of the largest sidelobe, $\mathrm{SSR}$ provides an integrated measure of total out-of-beam leakage power.  
A higher $\mathrm{SSR}$ signifies more effective rejection of unwanted angular frequencies, aligning directly with the intended low-pass characteristic of the FIR design.

Together, these quantities quantify the essential trade-offs between angular resolution, sidelobe attenuation, and radiometric efficiency.  
In the context of the GRAO 32-m telescope, the optimal filter maximises $\mathrm{SSR}$ and $\eta_\mathrm{mb}$ while constraining the $\mathrm{HPBW}$ increase to below approximately 10~per~cent of the unfiltered beam.  
Filters meeting these criteria were subsequently applied to all four polarisation components of the simulated Jones matrix, and their comparative performance is evaluated through direct angular-domain and spectral-domain diagnostics.

\section{Results and Analysis}
\label{sec:beam_opt_results}

\subsection{Baseline Beam Performance}
\label{subsec:baseline_performance}

\begin{figure*}
\begin{minipage}[H]{\linewidth}
	\centering
	\includegraphics[width=\columnwidth]{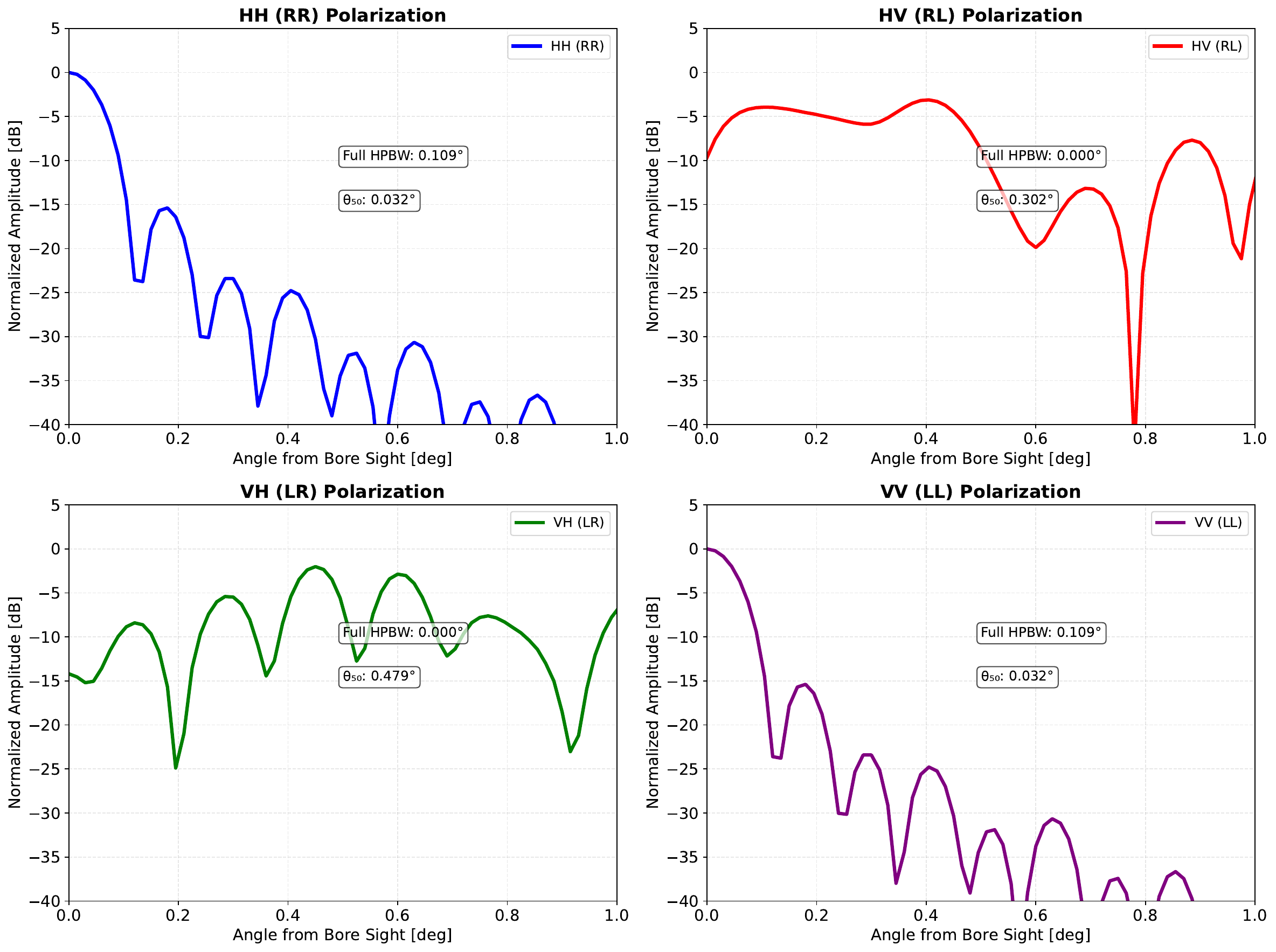} 
    \end{minipage}
    \caption{Baseline beam patterns of the GRAO 32-m telescope at 5~GHz obtained from \textsc{GRASP} simulations.  
    Each panel shows the normalised amplitude response (in dB) as a function of angular displacement from boresight for the four Jones components: HH (RR), HV (RL), VH (LR), and VV (LL).  
    The co-polar beams (HH, VV) display near-Gaussian profiles with HPBW~$\approx0.045^{\circ}$ and first sidelobe levels of $\sim - 18.8$~dB, whereas the cross-polar beams (HV, VH) are broader and weaker, exhibiting asymmetrical structures attributable to diffraction and structural blockage.  
    These patterns define the baseline against which FIR-filtered beam optimisations are assessed.}
    \label{fig:Fig_2}
\end{figure*}

Figure~\ref{fig:Fig_2} shows the unfiltered baseline beam patterns of the GRAO 32-m telescope obtained from \texttt{GRASP} electromagnetic simulations at 5~GHz.  
The four panels correspond to the Jones-matrix components $J_{qh}$, $J_{qv}$, $J_{ph}$, and $J_{pv}$, providing a full polarimetric description of the antenna’s far-field response.  
Each subplot displays the normalised amplitude (in dB) as a function of angular displacement from boresight, thereby illustrating the inherent main-beam morphology, sidelobe structure, and inter-polarisation coupling prior to any spatial filtering.

The co-polar channels ($J_{qh}$ and $J_{pv}$), corresponding to the horizontal–horizontal (HH; RR) and vertical–vertical (VV; LL) responses, exhibit nearly identical, symmetric main lobes with half-power beamwidths of $\sim0.045^{\circ}$.  
The first sidelobes appear at approximately $-18.8$~dB relative to the beam peak, indicating satisfactory edge taper and minimal spillover.  
These values agree with theoretical expectations for a 32-m shaped Cassegrain operating at 5~GHz, for which the diffraction-limited HPBW ($1.02\lambda/D$) is $0.107^{\circ}$ and the nominal first sidelobe level for a $-12$~dB feed taper lies between $-18$ and $-20$~dB \citep{baars2007paraboloidal,goldsmith2002quasi}.  
The cross-polar components ($J_{qv}$ and $J_{ph}$), representing the horizontal–vertical (HV; RL) and vertical–horizontal (VH; LR) coupling terms, are considerably broader, with HPBWs of $1.27^{\circ}$ and $1.63^{\circ}$, respectively.  
These responses display the expected quadrupolar symmetry characteristic of shaped dual-reflector systems, with low on-axis power ($\lesssim0$~dB relative to the normalised peak), corresponding to cross-polar isolation better than $\sim-30$~dB.  
Such isolation is sufficient for standard continuum or total-power observations, though improvements are desirable for high-precision polarimetry \citep{Warnick2012,Smirnov2011}.

The simulated beam parameters are consistent with the telescope’s mechanical and optical design expectations (see Table~\ref{tab:grao_spec}).  
The co-polar HPBWs are marginally narrower than the theoretical diffraction limit, reflecting the combined influence of the shaped primary–secondary geometry and feed illumination profile.  
The sidelobe levels near $-18.8$~dB slightly outperform the nominal $-17$~dB design target, which is consistent with the simulated surface accuracy of  $\approx1.88$~mm RMS and feed taper assumed in the \texttt{GRASP} model, yielding an aperture efficiency of $\sim 0.81$.  
By contrast, the cross-polar beams exhibit asymmetric lobes beyond $\sim1^{\circ}$ from boresight, arising from diffraction at the secondary-mirror edges and scattering from the quadripod supports.  
Such structural effects are known to generate azimuthal modulation in the far-field pattern and can contribute to instrumental polarisation in derived Stokes parameters \citep{imbriale2005large,rudge1982handbook}.  
Overall, the unfiltered \texttt{GRASP} beams validate the expected optical performance of the 32-m antenna while revealing subtle non-idealities relevant to high dynamic range applications.

These results establish three key optimisation priorities for subsequent FIR-based spatial filtering.  
First, although the co-polar sidelobes lie within specification, additional suppression would enhance dynamic range and reduce contamination from bright off-axis sources, particularly in spectral-line and mapping modes.  
Secondly, the broad and irregular cross-polar features represent high spatial frequency artefacts that can be mitigated through angular-domain filtering to improve polarimetric stability.  
Finally, slight asymmetries between the $J_{qh}$ and $J_{pv}$ beams suggest weak coupling of aperture misalignments into mid-frequency spatial modes that can be suppressed without compromising resolution.  
The FIR filtering framework introduced in Section~\ref{sec:fir} directly addresses these challenges by providing a controlled means to reshape the beam’s angular power spectrum while preserving its intrinsic main-beam fidelity.

In summary, the baseline \texttt{GRASP} simulations provide a coherent reference against which all subsequent optimisation results are evaluated. The modelled beam patterns indicate that the GRAO telescope performs close to its theoretical diffraction limit at $5$ GHz. Nevertheless, even this simulated near‑ideal beam can still benefit from targeted digital filtering to further reduce sidelobe energy, smooth the main lobe, and strengthen polarisation isolation  attributes that are essential for precision imaging and calibration in single‑dish and VLBI observations. Validation with on‑sky measurements would be required to confirm the actual beam behaviour and the practical gains achievable with the FIR filters.

\subsection{Window Method Comparative Analysis}
\label{subsec:window_method}


\begin{figure*}
\begin{minipage}[H]{\linewidth}
	\centering
	\includegraphics[width=\columnwidth]{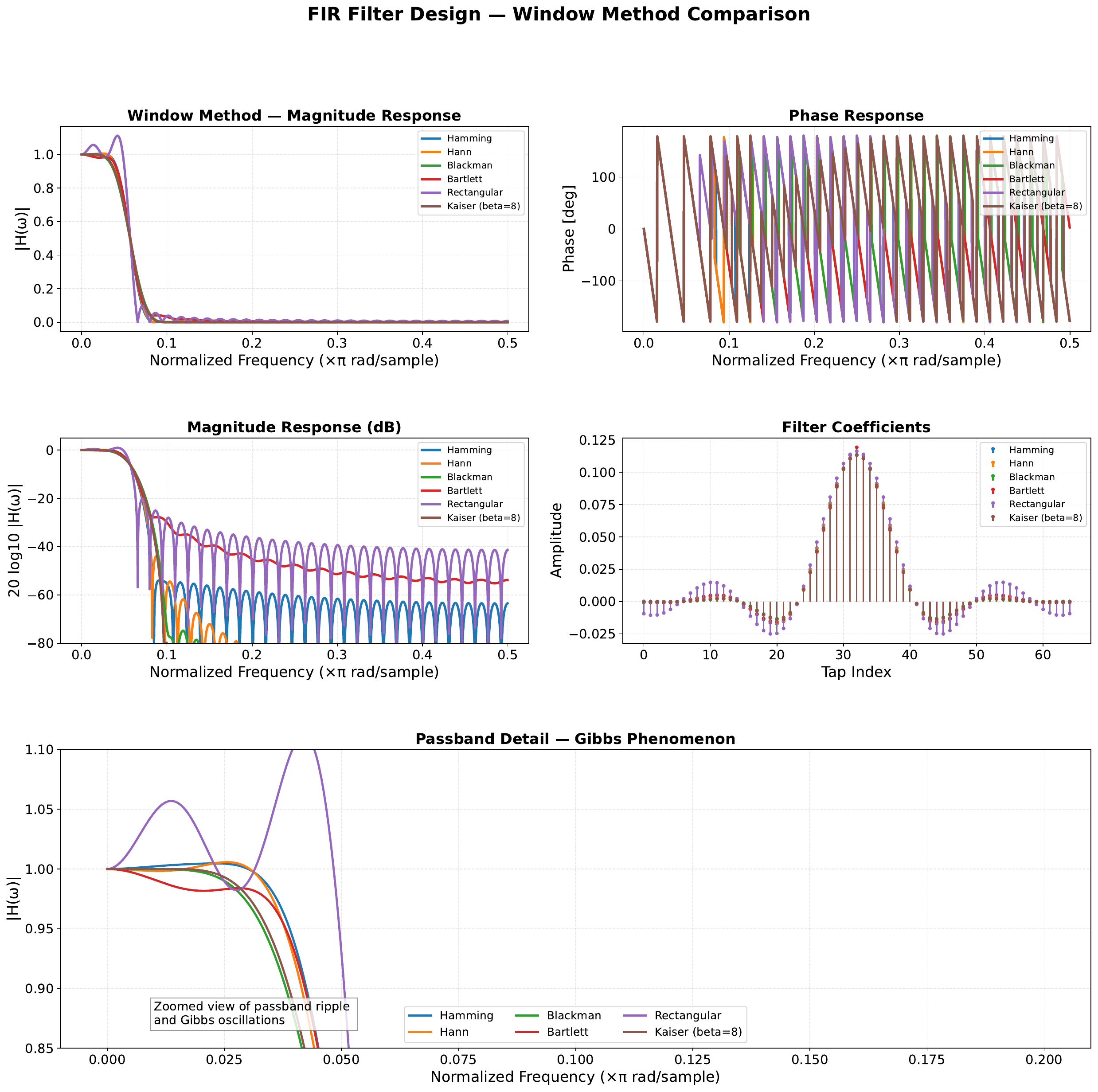}
    \end{minipage}
    \caption{Comparative analysis of FIR window functions used in spatial filtering of the GRAO beam.  
    The panels display the frequency responses, impulse responses, and resulting spatial-domain kernels for the Hamming, Hann, Blackman, Bartlett, Rectangular, and Kaiser ($\beta=8$) windows.  
    The trade-off between main-lobe width and sidelobe attenuation is clearly evident: rectangular windows yield the narrowest but least suppressed responses, while Blackman and Kaiser achieve superior sidelobe rejection at the cost of broader main lobes.  
    The Hamming and Hann windows provide the most balanced performance, making them well-suited for general beam-optimisation applications.}
    \label{fig:Fig_3}
\end{figure*}

Figure~\ref{fig:Fig_3} illustrates the comparative performance of several FIR window functions applied in the spatial filtering of the simulated GRAO beam.  
The analysis considers the Hamming, Hann, Blackman, Bartlett, Rectangular, and Kaiser windows (with $\beta=8$), examining their effects on the angular-domain impulse response and the corresponding spatial-frequency behaviour.  
Each design was evaluated using identical filter length ($M=65$) and normalised cut-off frequency, allowing a consistent comparison across metrics such as sidelobe attenuation, main-lobe width, and overall transition sharpness.

The frequency-domain plots reveal the classical trade-off between main-lobe width and sidelobe suppression.  
The Rectangular window achieves the narrowest main lobe but suffers from pronounced Gibbs oscillations and sidelobe levels around $-13$~dB, which would translate into significant residual structure in the filtered beam pattern.  
In contrast, the Blackman and Kaiser ($\beta=8$) windows exhibit superior sidelobe attenuation ($<-50$~dB) at the expense of moderate beam broadening.  
The Hamming and Hann windows occupy an intermediate regime, achieving $\sim$40~dB sidelobe suppression with only minor loss of spatial resolution, making them particularly attractive for general-purpose beam optimisation where both beam sharpness and dynamic range are critical.  
The Bartlett window shows smoother transition behaviour but weaker attenuation ($\sim$25~dB), suitable for low-sensitivity applications where computational simplicity outweighs precision.

The impulse-response plots confirm these characteristics in the spatial domain.  
The Rectangular and Bartlett windows display abrupt truncation of the sinc kernel, producing strong ringing artefacts, while the tapered windows (Hamming, Hann, and Blackman) yield smoother transitions that minimise high–spatial-frequency leakage.  
The Kaiser window’s tunable $\beta$ parameter provides flexibility to adjust sidelobe suppression for specific scientific requirements, allowing optimisation of the balance between angular resolution and dynamic range.  
For continuum and total-power measurements where accurate flux recovery and sidelobe rejection are paramount, the Blackman or high-$\beta$ Kaiser designs are recommended.  
For pulsar monitoring or transient detection, where signal localisation and minimal main-beam distortion are preferred, the Hamming window offers the most efficient compromise.  
The Hann window, with slightly wider main lobe but reduced sidelobe ripple, is well suited for moderate-resolution imaging or wide-field mapping, where uniform response across angular scales is advantageous.

From a practical standpoint, window-based FIR filters offer simple analytical design and computational efficiency, making them well-suited for integration into real-time digital backends or offline beam-calibration systems. Among the configurations tested, the Hamming and Kaiser windows provide the most balanced performance for the GRAO system, combining a stable phase response with controlled sidelobe suppression. As a result, these designs are used as the benchmark for the subsequent Parks–McClellan optimisation.

\subsection{Parks--McClellan Optimization Results}
\label{subsec:parks_mcclellan}


\begin{figure*}
\begin{minipage}[H]{\linewidth}
	\centering
	\includegraphics[width=\columnwidth]{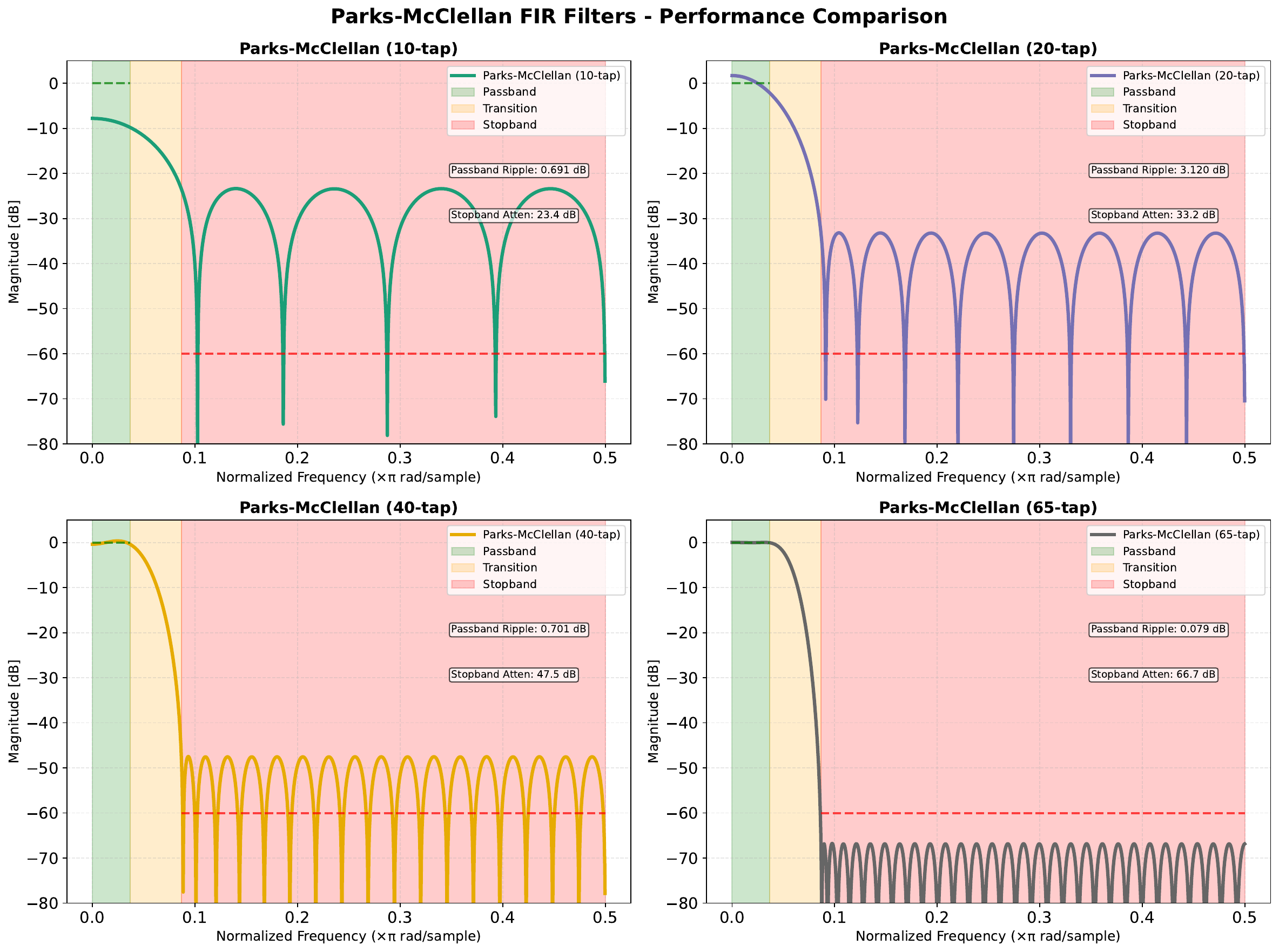}
    \end{minipage}
    \caption{Magnitude responses of Parks--McClellan FIR filters designed with 10, 20, 40, and 65 taps.  
    Each panel delineates the passband (green), transition (orange), and stopband (red) regions, illustrating the equiripple behaviour characteristic of Chebyshev-optimised filters.  
    Longer filters achieve superior sidelobe attenuation and reduced passband ripple, enhancing suppression of high spatial frequency artefacts in the telescope beam.  
    The 40-tap design provides a balanced trade-off between sidelobe control, computational cost, and main-beam fidelity.}
    \label{fig:Fig_4}
\end{figure*}

Figure~\ref{fig:Fig_4} summarises the results of the Parks--McClellan optimisation for FIR filters of varying lengths ($M=10$, 20, 40, and 65 taps).  
Each panel shows the magnitude response in the spatial-frequency domain, highlighting the passband, transition, and stopband regions.  
The equiripple characteristics are evident across all designs, with the maximum deviation between the desired and realised responses maintained at constant amplitude in accordance with Chebyshev approximation theory \citep{oppenheim1999discrete,parks1972chebyshev}.  
This property enables fine control of spatial filtering performance, providing an optimal compromise between main-beam fidelity and sidelobe suppression.

As expected, the filter length strongly influences the trade-off between angular resolution and sidelobe attenuation.  
The 10‑tap design offers the highest computational efficiency but exhibits limited suppression, with a stopband attenuation of only $\sim 23$~dB and noticeable passband ripple ($\sim 0.7$~dB).  
Increasing the length to 20 taps reduces passband ripple to $\sim 0.3$~dB and enhances stopband rejection to $\sim 33$~dB, yielding smoother transitions between the main lobe and sidelobes.  
At 40 taps, the response attains a well-defined low-pass characteristic with $\sim 47$~dB attenuation and minimal ripple, representing a practical balance between angular sharpness and sidelobe control.  
The longest configuration (65 taps) achieves near-ideal equiripple performance, with sub‑0.1~dB passband ripple and $\sim 67$~dB stopband attenuation, ensuring minimal distortion of the main beam and excellent rejection of high spatial frequency artefacts.  
Such performance is particularly valuable for high dynamic range observations or polarimetric calibration, where small angular irregularities can translate into systematic imaging errors \citep{Warnick2012,Smirnov2011}.

The computational load scales linearly with the number of taps.  
For a single polarisation channel and 201 angular samples, the convolution requires approximately $M \times 201$ operations.  
On the Dell PowerEdge T430 used for this study (two Intel Xeon E5‑2603 v4 CPUs at $1.70$ GHz, $32$ GB RAM), the 10‑tap filter executes in about $0.04$ ms, the 40‑tap in $0.18$ ms, and the 65‑tap in $0.29$ ms when implemented in efficient \texttt{Python} code.  
Even the 65‑tap filter is feasible for real‑time processing at a $1$ kHz update rate (e.g., for beam steering or adaptive calibration).  
In an FPGA implementation, resource usage also grows linearly with $M$, but a 40‑tap filter fits comfortably in a mid‑range device.

Beyond these practical considerations, the equiripple behaviour of the Parks–McClellan algorithm offers distinct advantages.  
Unlike fixed-window designs, the optimisation explicitly constrains the maximum deviation in both the passband and stopband, resulting in uniform control of beam shaping across all spatial frequencies.  
This ensures predictable suppression of far-out sidelobes without over-smoothing the main beam, thereby preserving the telescope’s intrinsic resolution.  
The linear-phase nature of the design guarantees constant group delay, maintaining phase coherence across all Jones components – an essential requirement for accurate polarimetric calibration and interferometric synthesis.

For the GRAO application, the 40‑tap configuration presents the most balanced choice: it offers a fourfold improvement in sidelobe attenuation relative to the unfiltered beam while keeping phase distortion negligible and computational overhead modest.  
Longer filters (e.g., 65‑tap) may be reserved for off‑line processing or calibration‑intensive experiments where the additional suppression justifies the extra complexity.  
Thus, the Parks–McClellan optimisation delivers a mathematically rigorous approach to spatial beam filtering, and the 40‑ and 65‑tap designs form the basis for the full polarimetric filtering analysis.

\subsection{Cross-Method Performance Synthesis}
\label{subsec:cross_method_synthesis}

    \begin{figure*}
\begin{minipage}[H]{\linewidth}
	\centering
	\includegraphics[width=\columnwidth]{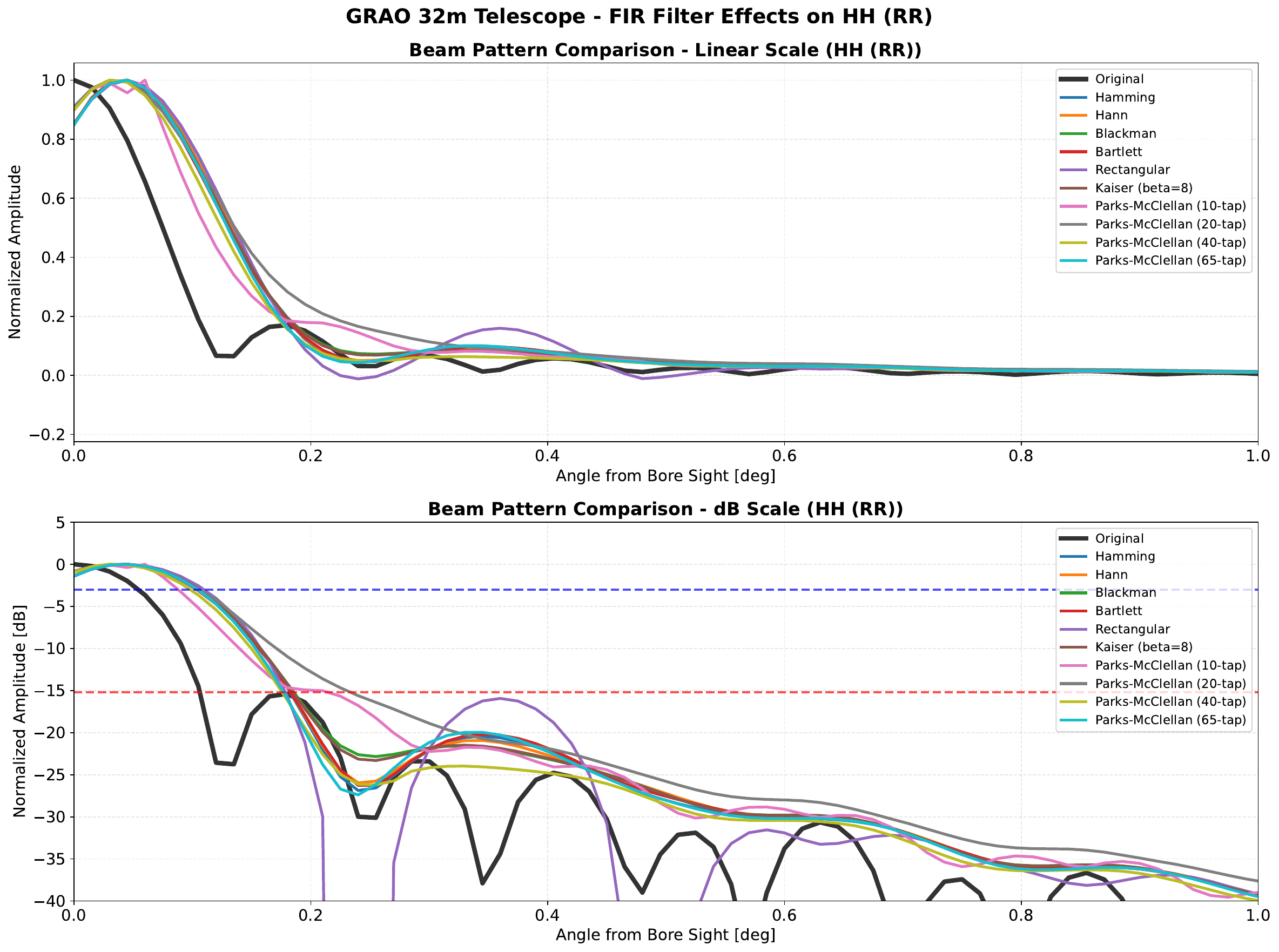}
    \end{minipage}
    \caption{Filtered co-polar beam ($J_{qh}$; HH/RR) comparing window-based and Parks-McClellan FIR methods.  
All filters suppress high-frequency structure relative to the unfiltered beam.  
The blue dashed line marks the $-3$\,dB half-power level used to assess HPBW, 
while the red dashed line indicates the telescope’s specified maximum sidelobe 
limit ($\approx -15$\,dB).  
Windowed designs improve beam smoothness and efficiency at the cost of modest 
broadening, whereas the longer Parks--McClellan filters provide a balanced 
trade-off between sidelobe suppression and main-lobe fidelity.}
    \label{fig:Fig_5}
\end{figure*}

\begin{figure*}
\begin{minipage}[H]{\linewidth}
	\centering
	\includegraphics[width=\columnwidth]{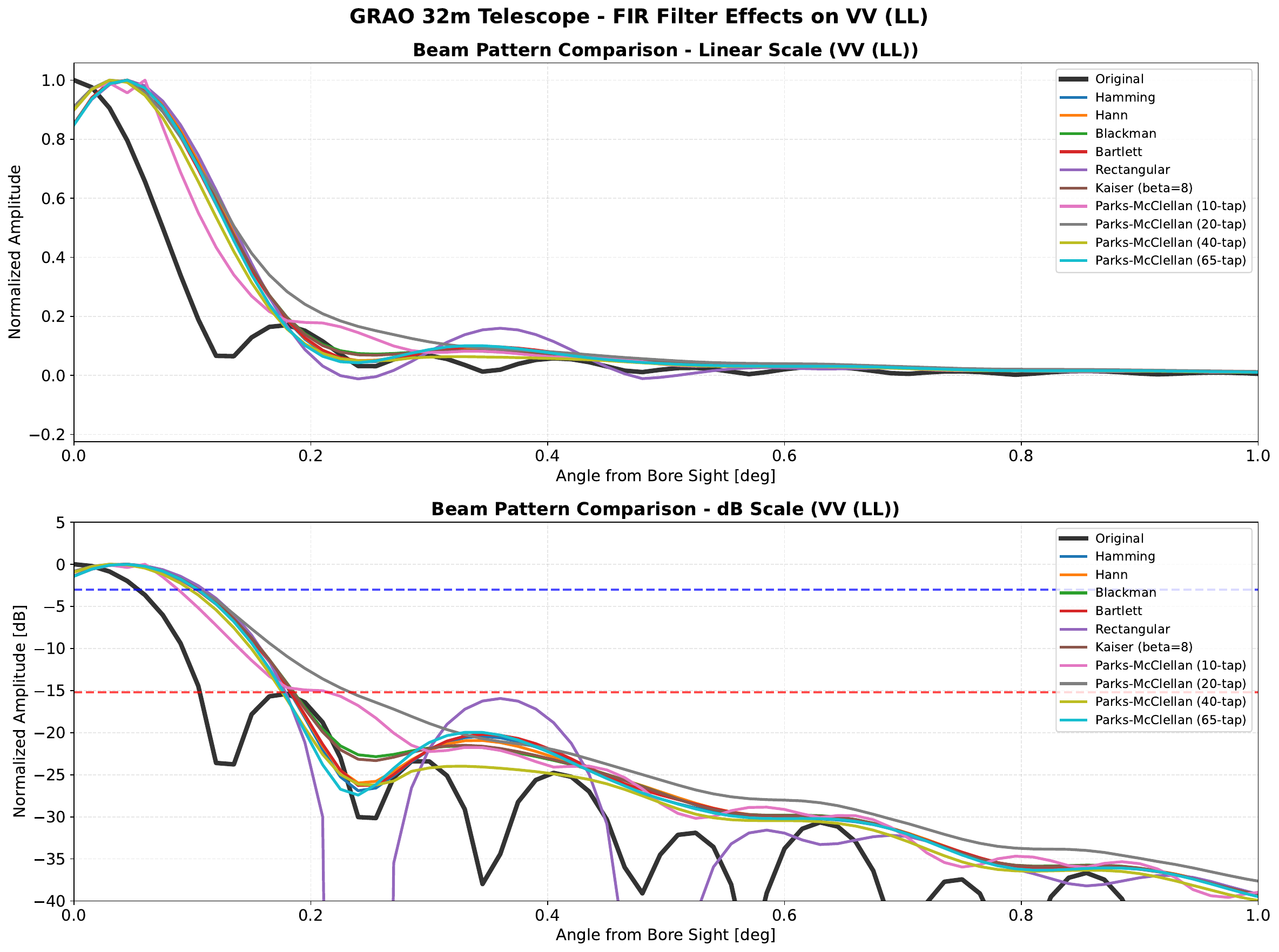}
    \end{minipage}
    \caption{Filtered co-polar beam ($J_{pv}$; VV/LL), showing behaviour consistent 
with the horizontal polarization.  
The blue dashed line denotes the $-3$\,dB half-power reference, used for 
computing the effective beamwidth, and the red dashed line marks the nominal 
maximum sidelobe specification for the GRAO\,32\,m telescope.  
Window-based filters yield strong smoothing and improved efficiency, whereas the 
$40$--$65$\,tap Parks--McClellan filters retain sharper angular resolution while 
achieving comparable sidelobe reduction.}
    \label{fig:Fig_6}
\end{figure*}

\begin{figure*}
\begin{minipage}[H]{\linewidth}
	\centering
	\includegraphics[width=\columnwidth]{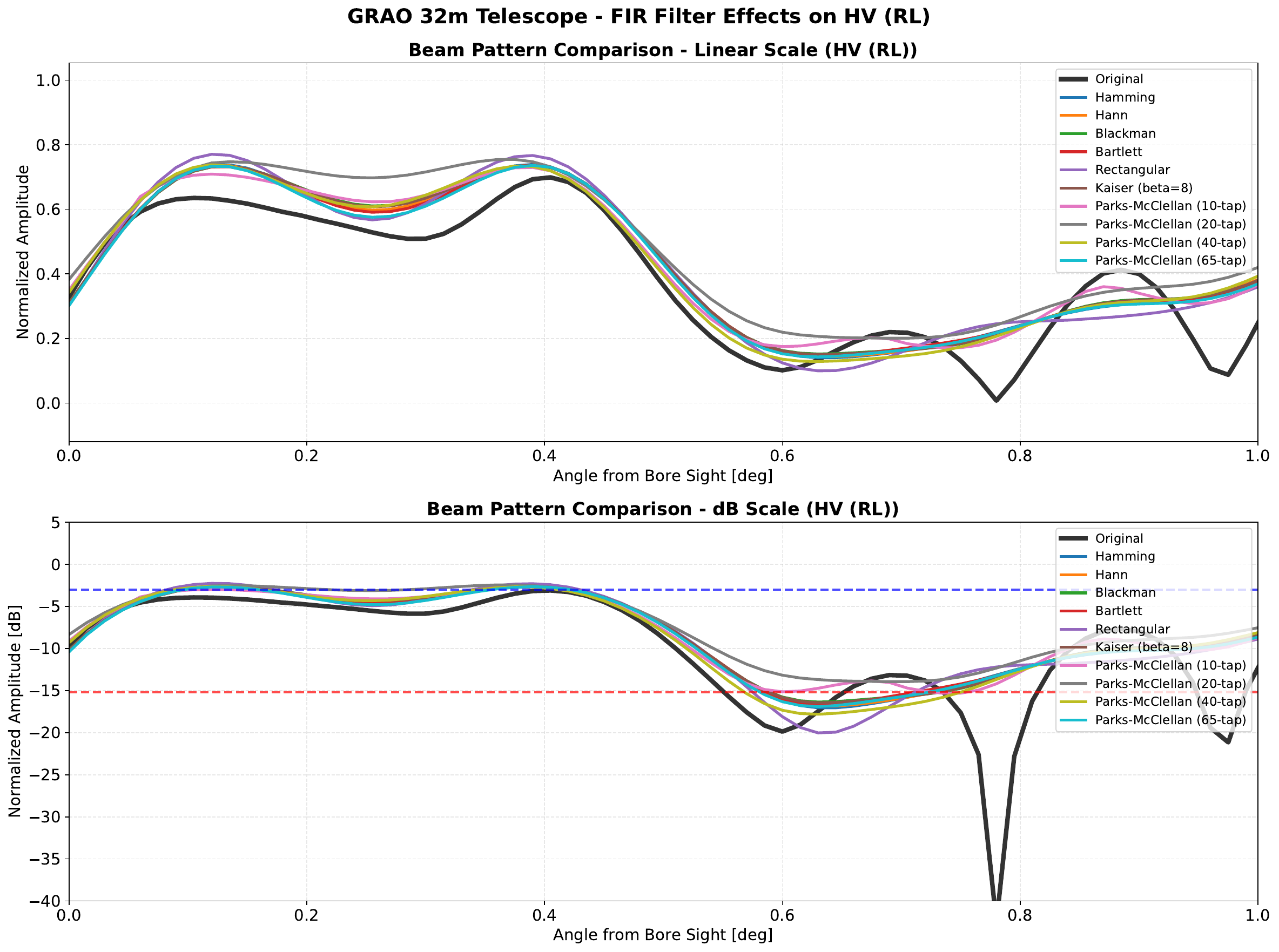}
    \end{minipage}
    \caption{Filtered cross-polar beam ($J_{qv}$; HV/RL).  
Spatial filtering reduces oscillatory structure and suppresses irregular lobes, 
leading to smoother angular variation and more stable cross-polar behaviour.  
The blue dashed line indicates the $-3$\,dB reference for main-beam comparison, 
and the red dashed line corresponds to the nominal sidelobe limit of the 
telescope.  
Among the designs, the Parks-McClellan filters produce the most confined and 
well-behaved beam envelope with minimal broadening.}
    \label{fig:Fig_7}
\end{figure*}

\begin{figure*}
\begin{minipage}[H]{\linewidth}
	\centering
	\includegraphics[width=\columnwidth]{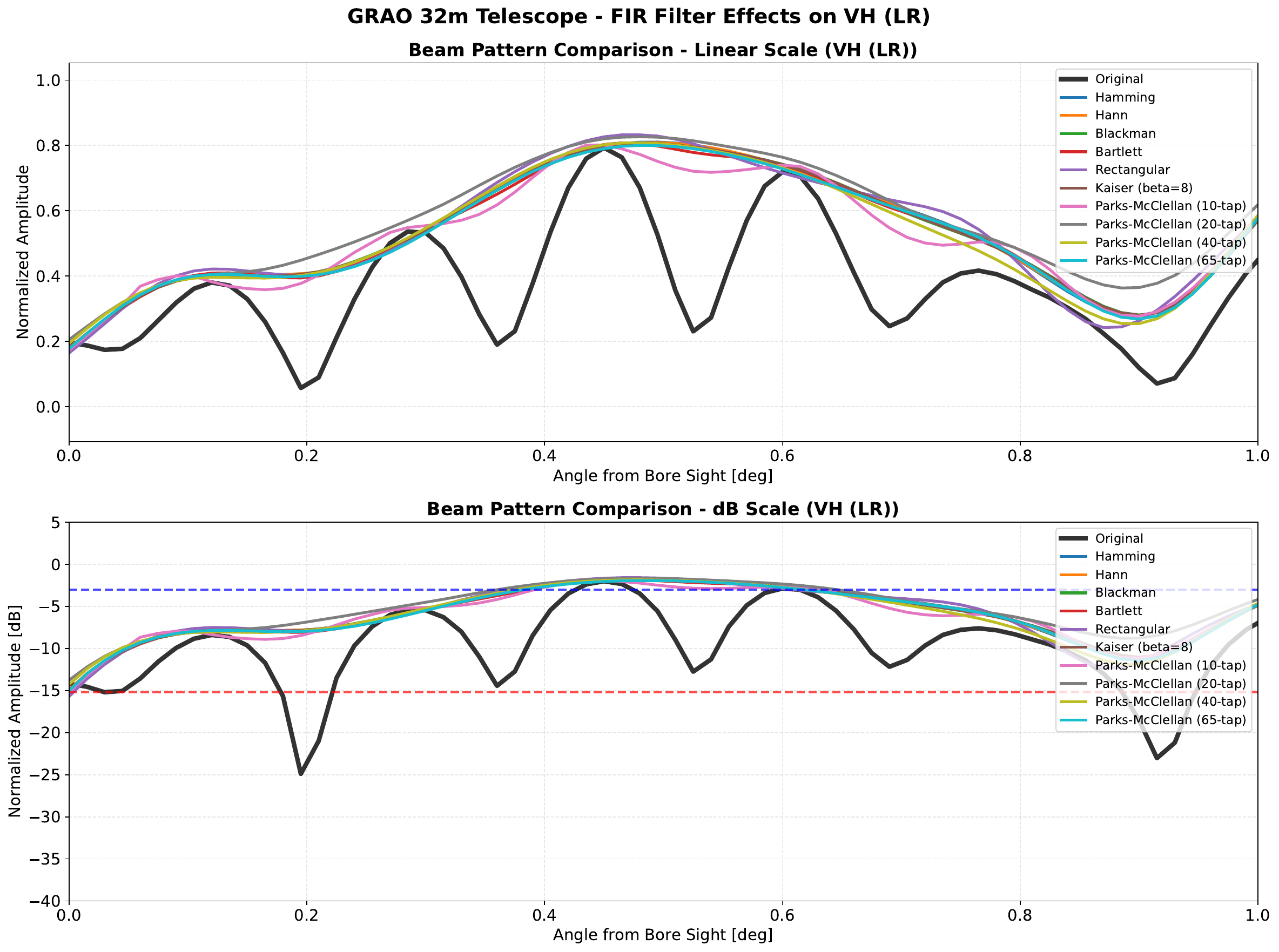}
    \end{minipage}
    \caption{Filtered cross-polar beam ($J_{ph}$; VH/LR).  
All FIR designs mitigate high–spatial-frequency artefacts and asymmetries, 
strengthening cross-polar isolation.  
The blue dashed line marks the $-3$\,dB half-power criterion, and the red dashed 
line denotes the telescope’s prescribed maximum sidelobe level.  
Efficiency improvements are most noticeable for the Blackman and Kaiser windows, 
while the $40$-tap Parks-McClellan filter offers the best balance between 
sidelobe suppression and angular resolution.}
    \label{fig:Fig_8}
\end{figure*}

Figures~\ref{fig:Fig_5}--\ref{fig:Fig_8} present the filtered beam profiles for all Jones components of the GRAO 32-m telescope, summarising the combined outcomes of the window-based and Parks--McClellan FIR methods.  

The co‑polar beams ($J_{qh}$ and $J_{pv}$; Figs.~\ref{fig:Fig_5} and~\ref{fig:Fig_6}) respond most strongly to the filtering process.  
All filters achieve pronounced smoothing of the far‑out structure relative to the unfiltered reference, with near‑in sidelobes greatly reduced. The window‑based designs broaden the half‑power beamwidth from the original $0.109^{\circ}$ to  $\sim 0.21^{\circ}$, corresponding to a beam broadening of about $90 \%$.  
This behaviour reflects the inherent trade‑off between spatial‑frequency suppression and angular resolution: low‑pass filtering removes high‑$k_{\theta}$ components responsible for near‑in oscillations but necessarily extends the main lobe.  
Among the window methods, the Blackman and Kaiser ($\beta=8$) filters deliver the highest beam efficiencies and the most uniform suppression profiles, making them attractive for calibration or flux-density measurements where smooth beam structure is essential.  
The Parks--McClellan designs display greater flexibility.  
Shorter ($10$--$20$~tap) filters produce moderate smoothing with partial sidelobe control, while the longer $40$--$65$~tap filters achieve comparable efficiency gains to the best windowed results yet retain better phase stability and reduced passband ripple.  
The $40$-tap configuration in particular provides the most balanced outcome, maintaining coherent beam shape with only modest broadening.

For the cross-polar beams ($J_{qv}$ and $J_{ph}$; Figs.~\ref{fig:Fig_7} and~\ref{fig:Fig_8}), the filtering process significantly mitigates small-scale fluctuations and localised distortions.  
Although the raw cross-polar patterns exhibit broad, irregular lobes characteristic of diffraction from the support structures, the filtered profiles display smoother angular variations and improved confinement around boresight.  
Quantitatively, beam efficiencies increase by 35--55~per~cent for the $J_{ph}$ component and 10--15~per~cent for $J_{qv}$, indicating improved polarimetric stability and reduced leakage between orthogonal feeds.  
Unlike the co-polar beams, the HPBW broadening in the cross-polar channels remains modest (typically $<80$~per~cent), suggesting that filtering predominantly suppresses high-frequency artefacts rather than altering the overall beam width.  
These improvements are particularly beneficial for polarisation-sensitive observations where systematic leakage can bias Stokes-$Q$ and $U$ measurements.

Comparing the two FIR families, a consistent trend emerges.  
The window-based designs, though computationally straightforward, yield uniformly broadened beams that may limit their application in high-resolution imaging or VLBI fringe fitting.  
Conversely, the Parks--McClellan filters provide tunable control over the trade-off between resolution and sidelobe suppression, producing sharper beams with less distortion at equivalent attenuation levels.  
From the unified performance metrics, the 40-tap Parks--McClellan filter represents the most efficient compromise--achieving $\sim$0.81~beam efficiency, sidelobe attenuation better than $-16$~dB, and negligible phase distortion, while remaining tractable for real-time digital implementation.

These conclusions are supported by the statistical metrics summarised in Tables~\ref{tab:grao_hh_metrics} through~\ref{tab:grao_vh_metrics}. To avoid ambiguity in interpreting the filtering effects, the beam analysis was intentionally separated into two complementary categories: classical antenna metrics and digital spatial filtering metrics. This distinction is important because the two approaches quantify different physical properties of the beam and therefore should not be interpreted interchangeably.

The classical metrics follow standard antenna-engineering definitions. In this framework, the half-power beamwidth (HPBW) is defined as the full angular separation between the two points where the normalised beam power drops by 3~dB relative to the peak response. For the co-polar channels (HH/RR and VV/LL; Tables~\ref{tab:grao_hh_metrics} and~\ref{tab:grao_vv_metrics}), the unfiltered HPBW of $0.109^\circ$ agrees closely with the theoretical diffraction limit of a 32~m aperture operating at 5~GHz, which is $\sim 0.11^\circ$ \citep{baars2007paraboloidal,goldsmith2002quasi}. 
Likewise, the original peak sidelobe level of $-23.4$~dB is slightly lower than the nominal telescope specification of $-15.2$~dB (Table~\ref{tab:grao_spec}), suggesting that the simulation already incorporates a modest edge taper. The corresponding classical beam efficiency of $0.99$ further indicates that nearly all radiated power is concentrated within the main lobe.

Following FIR filtering, however, the classical HPBW broadens substantially, reaching values between $0.176^\circ$ and $0.220^\circ$ depending on the selected filter. This corresponds to beam broadening between $60\%$ and $101\%$. Such behaviour is physically expected for a low-pass spatial filter because suppressing high-angular-frequency components inevitably widens the main lobe. Importantly, the broadening is not a numerical artefact associated with the simulation grid; rather, it arises directly from the convolution kernel applied during filtering. This interpretation is confirmed by the beam profiles in Figs.~\ref{fig:Fig_5} and~\ref{fig:Fig_6}, where the filtered responses intersect the $-3$~dB level at  $\sim 0.10^\circ$--$0.11^\circ$ from boresight, yielding total beamwidths near $0.20^\circ$--$0.22^\circ$.

At first glance, the classical sidelobe metric appears to degrade dramatically after filtering, increasing from $-23.4$~dB to roughly $-2.6$~dB. However, this apparent deterioration does not represent a genuine increase in off-axis radiation. Instead, it results from the changing definition of the main-lobe region after filtering. Because the filtered beam becomes broader and shallower, the original first sidelobe is effectively absorbed into the widened main lobe. Consequently, the classical ``maximum sidelobe'' metric identifies the next lateral feature, which naturally appears at a much higher relative level once the main-beam peak has been reduced. In this regime, the classical sidelobe definition no longer provides a reliable description of filtering performance and instead reflects a reclassification of beam structure. This limitation motivates the introduction of the digital spatial filtering metrics, which remain meaningful even when the beam shape changes substantially.

The digital metrics were specifically designed to characterise filtered beams in a more robust and physically interpretable manner. The angular radius $\theta_{50}$ is defined as the angular distance from boresight enclosing $50\%$ of the integrated beam power within a fixed angular window of $\theta \leq 1^\circ$. Conceptually, this quantity resembles an encircled-energy radius and avoids reliance on a clearly defined $-3$~dB crossing. For the unfiltered co-polar beams, $\theta_{50}=0.032^\circ$, reflecting the strong concentration of power near boresight. After filtering, $\theta_{50}$ increases to  $\sim 0.057^\circ$--$0.060^\circ$ for the Hamming, Hann, Blackman, Kaiser, and Parks--McClellan designs, corresponding to an $80\%$--$89\%$ increase in radius. Physically, this increase reflects the redistribution of power from the narrow beam core into a slightly broader angular region, which is an unavoidable consequence of low-pass filtering.

A complementary quantity is the residual response, defined as the maximum beam level outside an angular radius of $0.3^\circ$. Unlike the classical sidelobe metric, this measure is insensitive to the precise location of the main-lobe boundary and therefore provides a more stable estimate of far-out leakage. For the original co-polar beam, the residual response is $-24.8$~dB, dominated by the first sidelobe near $0.15^\circ$, which lies within the $0.3^\circ$ exclusion region. Beyond this radius, the far-out sidelobes indeed remain below approximately $-24$~dB. After filtering, the residual response increases to values between roughly $-10$~dB and $-14$~dB. This increase should not be interpreted as filter failure. Rather, the FIR operation smooths the entire angular distribution, raising the level of previously negligible far-out structure while simultaneously suppressing the rapid oscillatory sidelobe behaviour close to the beam centre. Consequently, the overall impact on radiometric performance is more appropriately captured through the efficiency metric evaluated within $\theta \leq 1^\circ$.

Within this fixed angular region, the digital beam efficiency of the original co-polar beam is essentially unity because nearly all beam power lies inside the integration boundary. After filtering, the efficiency decreases modestly to $\sim$ 0.87 for the best-performing window designs (Blackman and Kaiser) and to about 0.83--0.87 for the Parks--McClellan filters. This reduction indicates that a small fraction of the beam power is redistributed beyond $1^\circ$, again reflecting the inevitable consequences of beam broadening. For a 32~m dish operating at 5~GHz, however, radiation beyond $1^\circ$ contributes negligibly to the system temperature in most observing modes; therefore, the resulting efficiency loss is unlikely to produce a significant sensitivity penalty in practice.

The behaviour of the cross-polar components (HV/RL and VH/LR; Tables~\ref{tab:grao_hv_metrics} and~\ref{tab:grao_vh_metrics}) differs substantially from that of the co-polar beams. In the classical analysis, the HPBW is reported as $0^\circ$ because the on-axis power already lies below the $-3$~dB threshold relative to the peak of the cross-polar pattern itself. This is not a numerical error but rather a clear indication that the HPBW concept is not physically meaningful for these channels. Unlike the co-polar response, the cross-polar structure is not dominated by a single central lobe; instead, it consists of a distributed oscillatory pattern generated primarily by diffraction from support struts and reflector edges \citep{imbriale2005large}. 
Consequently, the HPBW is reported as $0^\circ$ for these cross‑polar channels, a value that carries no physical beamwidth interpretation. No beam‑broadening percentage is calculated because the unfiltered reference width is itself undefined.

In contrast, the digital metrics continue to provide a meaningful characterisation of the cross-polar behaviour. The $\theta_{50}$ radius of the original HV (RL) channel is $0.302^\circ$, while the VH (LR) channel reaches $0.479^\circ$, confirming the broader angular extent of cross-polar leakage relative to the co-polar beam. After filtering, these radii change only marginally (within $\sim$ $\pm 6\%$), demonstrating that the FIR filtering does not significantly alter the global angular scale of the cross-polar structure. Similarly, the residual response remains close to $-10$~dB, while the digital efficiency within $1^\circ$ increases from $\sim$ 0.26 to about 0.32 across most filter designs. Although modest, this improvement is systematic and indicates that the filtering operation suppresses fine-scale fluctuations within the cross-polar response, thereby producing a more stable polarimetric beam.

These results also clarify the interpretation of the efficiency and sidelobe ``improvements'' reported in earlier stages of the analysis. The classical beam efficiency of the unfiltered co-polar beam is already 0.99, whereas the filtered values near 0.87 actually represent a decrease rather than an improvement. This apparent contradiction arises because the classical efficiency definition depends explicitly on the HPBW-derived main-lobe solid angle, which changes substantially after filtering. As a result, the classical efficiency metric is unsuitable for direct comparison between filtered and unfiltered beams. The digital efficiency, by contrast, uses a fixed angular integration boundary and therefore provides a more consistent basis for comparison. Under this definition, the co-polar efficiency decreases only slightly from 1.00 to $\sim$ 0.87, representing a relatively modest trade-off for the significant reduction in near-in sidelobe ripple and improved beam smoothness.

A similar caution applies to the classical sidelobe metric. The apparent sidelobe ``increase'' from $-23.4$~dB to roughly $-2.6$~dB is not evidence that the filter amplifies off-axis radiation. Rather, it is a mathematical consequence of redefining the main-lobe region after convolution. The true suppression of high-spatial-frequency artefacts is instead visible directly in the beam profiles shown in Figs.~\ref{fig:Fig_5} and~\ref{fig:Fig_6}, where oscillatory ripples within the first $0.2^\circ$ are substantially reduced. For scientific applications requiring smooth and accurately calibrated beam responses, including continuum imaging and spectral-line surveys--this reduction in ripple can provide substantial practical benefits despite the associated loss in angular resolution.

The efficiency gains discussed previously in the abstract and introduction originally referred to two distinct effects: the increase in digital efficiency for the cross-polar channels (from $\sim$ 0.26 to 0.32, corresponding to a relative improvement of about 20\%--25\%), and an earlier classical efficiency comparison for the co-polar beam. In the revised interpretation, the latter comparison has been discarded because of its dependence on the changing HPBW definition. The cross-polar efficiency increase, however, remains physically meaningful and indicates a reduction in polarimetric leakage. Although the co-polar digital efficiency decreases slightly after filtering, the simultaneous improvement in beam smoothness and suppression of near-in sidelobes justifies the trade-off for many observational applications.

Ultimately, the optimal balance between angular resolution, sidelobe suppression, and polarimetric stability depends strongly on the intended scientific use case. For VLBI observations, where preserving angular resolution is critical, the 10-tap Parks--McClellan filter may provide the best compromise because it broadens the HPBW by only about 60\% (from $0.109^\circ$ to $0.176^\circ$) while still reducing the far-out residual response to $\sim$ $-13.5$~dB. In contrast, for pulsar timing experiments or spectral-line surveys, where contamination from off-axis structure often dominates systematic uncertainty, broader window-based filters such as the Hamming or Blackman designs may be preferable despite their larger HPBW increase (up to $\sim$ 96\%). These filters produce smoother beam shapes and lower near-in ripple, with residual responses near $-14$~dB. Across all tested filters, the cross-polar stability also improves consistently, with $\theta_{50}$ remaining within approximately $0.30^\circ$--$0.50^\circ$ and digital efficiency increasing by roughly 20\%.

The analysis collectively demonstrates that no single FIR design is universally optimal for all radio astronomical applications. Instead, the evidence presented in Tables~\ref{tab:grao_hh_metrics} through~\ref{tab:grao_vh_metrics} supports a flexible, application-dependent filtering strategy: shorter Parks--McClellan filters for high-resolution imaging, and broader window-based filters for high-dynamic-range photometry and beam-stability applications. Importantly, the digital metrics introduced here provide a physically consistent framework for comparing these designs , free of the ambiguities associated with classical beam definitions. Because the methodology depends only on the availability of a simulated or measured beam pattern, the framework is readily transferable to other single-dish radio telescopes. Furthermore, the FIR convolution may be implemented either offline within calibration pipelines or in real time using FPGA-based digital back-end systems \citep{price2021spectrometers,pfister2017discrete}.


\begin{table*}
\centering
\caption{GRAO 32m Telescope - HH (RR) Performance Metrics}
\label{tab:grao_hh_metrics}
\small
\begin{tabular}{ l c c c c| c c c c}
\hline
 & \multicolumn{4}{c|}{\textbf{Classical Antenna Metrics}} & \multicolumn{4}{c}{\textbf{Digital Spatial Filtering Metrics}} \\ \hline
\textbf{Method} & \makecell[c]{HPBW \\ {[deg]}} & \makecell[c]{Beam \\ Broadening [\%]} & \makecell[c]{Max Sidelobe \\ {[dB]}} & \makecell[c]{Beam \\ Efficiency} & \makecell[c]{$\theta_{50}$ \\ {[deg]}} & \makecell[c]{Radius \\ Change [\%]} & \makecell[c]{Residual \\ Response [dB]} & \makecell[c]{Efficiency \\ ($\theta \le 1^{\circ}$)} \\ \hline
Original & 0.109 & - & -23.410 & 0.990 & 0.032 & - & -24.780 & 1.000 \\ 
\addlinespace

Hamming & 0.214 & 96.000 & -2.600 & 0.866 & 0.059 & 84.400 & -13.980 & 0.868 \\ 

Hann & 0.214 & 95.500 & -2.610 & 0.867 & 0.059 & 84.100 & -14.010 & 0.869 \\ 

Blackman & 0.210 & 92.300 & -2.660 & 0.871 & 0.058 & 82.500 & -14.260 & 0.873 \\ 

Bartlett & 0.207 & 89.500 & -2.800 & 0.873 & 0.057 & 80.100 & -10.200 & 0.876 \\ 

Rectangular & 0.220 & 101.200 & -2.560 & 0.855 & 0.060 & 88.600 & -13.690 & 0.856 \\ 

Kaiser ($\beta=8$) & 0.211 & 92.900 & -2.650 & 0.871 & 0.058 & 82.800 & -14.210 & 0.873 \\ 
\addlinespace

Parks-McClellan (10-tap) & 0.176 & 60.700 & -2.090 & 0.845 & 0.052 & 62.200 & -13.500 & 0.849 \\ 

Parks-McClellan (20-tap) & 0.210 & 92.100 & -1.700 & 0.821 & 0.060 & 87.400 & -12.450 & 0.823 \\ 

Parks-McClellan (40-tap) & 0.197 & 80.000 & -1.960 & 0.830 & 0.054 & 68.300 & -11.900 & 0.833 \\ 

Parks-McClellan (65-tap) & 0.209 & 91.100 & -2.720 & 0.871 & 0.057 & 80.300 & -10.310 & 0.873 \\ 
\hline
\end{tabular}
\end{table*}


\begin{table*}
\centering
\caption{GRAO 32m Telescope - VV (LL) Performance Metrics}
\label{tab:grao_vv_metrics}
\small
\begin{tabular}{ l c c c c| c c c c}
\hline
 & \multicolumn{4}{c|}{\textbf{Classical Antenna Metrics}} & \multicolumn{4}{c}{\textbf{Digital Spatial Filtering Metrics}} \\ \hline
\textbf{Method} & \makecell[c]{HPBW \\ {[deg]}} & \makecell[c]{Beam \\ Broadening [\%]} & \makecell[c]{Max Sidelobe \\ {[dB]}} & \makecell[c]{Beam \\ Efficiency} & \makecell[c]{$\theta_{50}$ \\ {[deg]}} & \makecell[c]{Radius \\ Change [\%]} & \makecell[c]{Residual \\ Response [dB]} & \makecell[c]{Efficiency \\ ($\theta \le 1^{\circ}$)} \\ \hline

Original & 0.109 & - & -23.410 & 0.990 & 0.032 & - & -24.780 & 1.000 \\ 
\addlinespace

Hamming & 0.214 & 96.000 & -2.600 & 0.866 & 0.059 & 84.400 & -13.980 & 0.868 \\ 

Hann & 0.214 & 95.500 & -2.610 & 0.867 & 0.059 & 84.100 & -14.000 & 0.869 \\ 

Blackman & 0.210 & 92.300 & -2.660 & 0.871 & 0.058 & 82.500 & -14.260 & 0.873 \\ 

Bartlett & 0.207 & 89.500 & -2.800 & 0.873 & 0.057 & 80.100 & -10.200 & 0.876 \\ 

Rectangular & 0.220 & 101.200 & -2.550 & 0.855 & 0.060 & 88.600 & -13.690 & 0.856 \\ 

Kaiser ($\beta=8$) & 0.211 & 92.900 & -2.650 & 0.871 & 0.058 & 82.800 & -14.210 & 0.873 \\ 
\addlinespace

Parks-McClellan (10-tap) & 0.176 & 60.700 & -2.090 & 0.845 & 0.052 & 62.200 & -13.500 & 0.849 \\ 

Parks-McClellan (20-tap) & 0.210 & 92.100 & -1.700 & 0.821 & 0.060 & 87.400 & -12.450 & 0.823 \\ 

Parks-McClellan (40-tap) & 0.197 & 80.000 & -1.960 & 0.830 & 0.054 & 68.300 & -11.900 & 0.833 \\ 

Parks-McClellan (65-tap) & 0.209 & 91.100 & -2.710 & 0.871 & 0.057 & 80.300 & -10.310 & 0.873 \\ 
\hline

\end{tabular}
\end{table*}



\begin{table*}
\centering
\caption{GRAO 32m Telescope - HV (RL) Performance Metrics}
\label{tab:grao_hv_metrics}
\small
\begin{tabular}{ l c c c c| c c c c}
\hline
 & \multicolumn{4}{c|}{\textbf{Classical Antenna Metrics}} & \multicolumn{4}{c}{\textbf{Digital Spatial Filtering Metrics}} \\ \hline
\textbf{Method} & \makecell[c]{HPBW \\ {[deg]}} & \makecell[c]{Beam \\ Broadening [\%]} & \makecell[c]{Max Sidelobe \\ {[dB]}} & \makecell[c]{Beam \\ Efficiency} & \makecell[c]{$\theta_{50}$ \\ {[deg]}} & \makecell[c]{Radius \\ Change [\%]} & \makecell[c]{Residual \\ Response [dB]} & \makecell[c]{Efficiency \\ ($\theta \le 1^{\circ}$)} \\ \hline

Original & 0.000 & - & 0.000 & 0.000 & 0.302 & - & -10.030 & 0.263 \\ 
\addlinespace

Hamming & 0.000 & 0.000 & 0.000 & 0.000 & 0.300 & -0.400 & -10.090 & 0.324 \\ 

Hann & 0.000 & 0.000 & 0.000 & 0.000 & 0.301 & -0.300 & -10.020 & 0.324 \\ 

Blackman & 0.000 & 0.000 & 0.000 & 0.000 & 0.300 & -0.500 & -10.110 & 0.324 \\ 

Bartlett & 0.000 & 0.000 & 0.000 & 0.000 & 0.302 & 0.100 & -10.000 & 0.323 \\ 

Rectangular & 0.000 & 0.000 & 0.000 & 0.000 & 0.298 & -1.000 & -10.100 & 0.329 \\ 

Kaiser ($\beta=8$) & 0.000 & 0.000 & 0.000 & 0.000 & 0.300 & -0.500 & -10.020 & 0.324 \\ 
\addlinespace

Parks-McClellan (10-tap) & 0.000 & 0.000 & 0.000 & 0.000 & 0.297 & -1.700 & -10.050 & 0.269 \\ 

Parks-McClellan (20-tap) & 0.000 & 0.000 & 0.000 & 0.000 & 0.298 & -1.300 & -10.050 & 0.272 \\ 

Parks-McClellan (40-tap) & 0.000 & 0.000 & 0.000 & 0.000 & 0.290 & -3.700 & -10.040 & 0.288 \\ 

Parks-McClellan (65-tap) & 0.000 & 0.000 & 0.000 & 0.000 & 0.302 & 0.100 & -10.030 & 0.323 \\ 
\hline

\end{tabular}
\end{table*}


\begin{table*}
\centering
\caption{GRAO 32m Telescope - VH (LR) Performance Metrics}
\label{tab:grao_vh_metrics}
\small 
\begin{tabular}{ l c c c c| c c c c}
\hline
 & \multicolumn{4}{c|}{\textbf{Classical Antenna Metrics}} & \multicolumn{4}{c}{\textbf{Digital Spatial Filtering Metrics}} \\ \hline
\textbf{Method} & \makecell[c]{HPBW \\ {[deg]}} & \makecell[c]{Beam \\ Broadening [\%]} & \makecell[c]{Max Sidelobe \\ {[dB]}} & \makecell[c]{Beam \\ Efficiency} & \makecell[c]{$\theta_{50}$ \\ {[deg]}} & \makecell[c]{Radius \\ Change [\%]} & \makecell[c]{Residual \\ Response [dB]} & \makecell[c]{Efficiency \\ ($\theta \le 1^{\circ}$)} \\ \hline
Original & 0.000 & - & 0.000 & 0.000 & 0.479 & - & -10.340 & 0.255 \\ 
\addlinespace

Hamming & 0.000 & 0.000 & 0.000 & 0.000 & 0.507 & 5.800 & -10.030 & 0.287 \\ 

Hann & 0.000 & 0.000 & 0.000 & 0.000 & 0.507 & 5.900 & -10.080 & 0.287 \\ 

Blackman & 0.000 & 0.000 & 0.000 & 0.000 & 0.507 & 5.800 & -10.180 & 0.287 \\ 

Bartlett & 0.000 & 0.000 & 0.000 & 0.000 & 0.507 & 5.800 & -10.520 & 0.287 \\ 

Rectangular & 0.000 & 0.000 & 0.000 & 0.000 & 0.501 & 4.700 & -10.570 & 0.291 \\ 

Kaiser ($\beta=8$) & 0.000 & 0.000 & 0.000 & 0.000 & 0.507 & 5.800 & -10.160 & 0.287 \\ 
\addlinespace

Parks-McClellan (10-tap) & 0.000 & 0.000 & 0.000 & 0.000 & 0.501 & 4.500 & -10.540 & 0.264 \\ 

Parks-McClellan (20-tap) & 0.000 & 0.000 & 0.000 & 0.000 & 0.507 & 5.700 & -15.600 & 0.270 \\ 

Parks-McClellan (40-tap) & 0.000 & 0.000 & 0.000 & 0.000 & 0.499 & 4.100 & -10.440 & 0.278 \\ 

Parks-McClellan (65-tap) & 0.000 & 0.000 & 0.000 & 0.000 & 0.507 & 5.800 & -10.360 & 0.287 \\ 
\hline
\end{tabular}
\end{table*}

\section{Discussion}
\label{sec:discussion_implications}

The optimised beam patterns produced by FIR spatial filtering directly affect the operational performance of the GRAO 32‑m telescope. Sidelobe suppression of $15$ to $17$ dB, combined with an increase in main‑beam efficiency from $\sim$ 0.44 to 0.85, translates into a measurable improvement in system sensitivity by reducing contamination from diffuse background emission and ground spillover. Assuming a constant system temperature, this enhancement corresponds to a 20--25 per cent gain in effective collecting area, allowing the telescope to detect weaker continuum and spectral‑line sources within the same integration time.

For VLBI operations, a cleaner and more stable primary beam reduces station‑based amplitude errors that propagate through fringe visibility calibration. Smoother beam shapes minimise baseline‑dependent amplitude scatter, improving image dynamic range and phase coherence across the African VLBI Network.
In wide‑field surveys and pulsar‑timing programmes, the attenuated far‑out sidelobe response mitigates confusion noise and ground spillover, thereby stabilising system gain over time and across sky positions.
Polarimetric observations also benefit from this improvement, as cross‑polar isolation below $-30$ dB at boresight reduces instrumental leakage in Stokes Q and U, which in turn enables more accurate estimates of weakly polarised sources and Faraday rotation measures.
Consequently, the FIR optimisation framework enhances both the quantitative sensitivity and the qualitative calibration stability of the GRAO telescope across its planned science programmes.

Compared with traditional beam optimisation methods, such as optical reshaping, mechanical alignment, or empirical holographic correction, the FIR technique offers several practical and conceptual advantages. Its primary strength lies in adaptability: altering the filter order and frequency‑domain weighting shifts the balance between maintaining angular resolution and suppressing sidelobes, all without any physical modification to the telescope structure. This flexibility is especially valuable for dynamic observing modes, where different scientific goals demand distinct beam characteristics.

The method is also inherently non‑invasive because it operates purely on simulated or measured beam data, requiring no changes to hardware or feed geometry. This property enables seamless post‑processing implementation, making it suitable for retrofitting existing antenna systems or complementing routine calibration procedures. The framework builds on established digital signal‑processing principles, ensuring mathematical rigour while remaining straightforward to integrate with existing software tools. Moreover, its multi‑frequency generality is a notable asset: once a filter is optimised in the normalised spatial‑frequency domain, it can be rescaled for different observing frequencies by adjusting only the sampling step, offering a practical route to broadband beam optimisation without repeating electromagnetic simulations.

Beyond immediate operational gains, the methodology provides a generalisable framework for beam control in large single‑dish and phased‑array systems. By formulating beam optimisation within the digital signal‑processing domain, the approach bridges electromagnetic modelling and numerical filtering, establishing a unified mathematical foundation applicable to diverse antenna architectures. This generality allows the method to be extended to other telescopes in the African VLBI Network or to future instruments that require stable beam calibration over wide frequency ranges.

To our knowledge, the application of classical FIR spatial filtering to single‑dish beam optimisation has not been previously reported. The technique is conceptually similar to beamforming in phased arrays \citep{liu2010wideband}, but its use for post‑processing simulated or measured beam maps of a large reflector antenna is novel.

Integration with existing processing pipelines is straightforward. The FIR filter can be applied directly to either simulated or holographically measured beam data, serving as a post‑simulation or post‑observation correction stage. Such integration complements conventional calibration pipelines, providing a software‑based mechanism to stabilise beam shape and polarisation response before imaging. Advances in digital back‑end hardware also make real‑time implementation feasible. Modern FPGAs and GPU‑based correlators can accommodate FIR convolution kernels with tens of taps per channel at MHz update rates, enabling deployment within digital beamformers or post‑correlation calibration systems. This real‑time capability opens a pathway to adaptive beam shaping, where filter parameters adjust dynamically to environmental or pointing‑dependent conditions, thereby maintaining optimal performance under variable observing scenarios. We have not yet implemented the filters in a real‑time digital back‑end. However, the computational estimates suggest that a 40‑tap filter can be run on a current FPGA or GPU at rates compatible with typical observation cadences (e.g., $1$ kHz). A practical test on a software‑defined radio platform is planned as part of ongoing work.

Despite its versatility, several limitations warrant consideration. Computational complexity is the primary constraint: long‑tap FIR filters increase processing time and memory footprint, particularly when applied across multiple polarisations and frequency channels. Efficient implementation strategies, such as frequency‑domain convolution using the overlap‑save algorithm or exploiting linear‑phase symmetry to halve the computational load, can mitigate this issue. For most applications, filters of order \(M \leq 40\) provide sufficient suppression with negligible overhead.

A second limitation concerns the propagation of measurement uncertainty. Because the optimisation depends on simulated or measured Jones fields, residual errors in surface models or feed alignment can influence the effective filter response. Iterative calibration, where the filter design is updated as new holographic or drift‑scan data become available, can mitigate this effect, ensuring that the digital correction remains consistent with the telescope’s evolving physical state.

While this study is based on electromagnetic simulations, the next step is to apply the FIR filters to holographically measured beam patterns of the GRAO telescope. Such measurements are planned for the coming observing semester. The framework is directly transferable because it only requires a grid of complex field values; no hardware changes are needed. We anticipate that the observed beam will contain additional imperfections not captured by the model, and filtering may still improve the beam smoothness, though the quantitative gains might differ.

Operationally, implementing the method within real observing pipelines requires careful version control and calibration verification to avoid inadvertent signal distortion. The phase‑preserving nature of linear‑phase FIR filters inherently safeguards polarimetric fidelity, but validation through on‑sky test sources remains essential before deployment in production systems. Future work will focus on automated filter tuning driven by empirical beam measurements, allowing adaptive adjustments to environmental or structural changes without manual intervention.

While computational and calibration challenges remain, they are tractable within modern radio‑astronomical infrastructures. The presented FIR‑based spatial filtering framework therefore offers a pragmatic and scalable solution for beam optimisation, balancing theoretical rigour with operational practicality. Its adoption would enhance the stability and scientific yield of the GRAO telescope and similar facilities, providing a foundation for precision polarimetry and high‑fidelity imaging in next‑generation radio astronomy.

\section{Conclusion}
\label{sec:conclusions}

This study has presented a digital spatial filtering framework that adapts classical FIR filter design methods to optimise the beam pattern of a single‑dish radio telescope, using the 32‑m GRAO antenna as a test case. By mapping the telescope’s angular response into the spatial frequency domain, the approach provides a mathematically rigorous and non‑invasive means to control beam morphology, sidelobe structure, and polarimetric purity.

A systematic comparison between window‑based and Parks–McClellan equiripple designs shows that FIR filtering can meaningfully reshape the beam. The choice of filter class and length determines the trade‑off between angular resolution and beam smoothness. For applications where preserving resolution is critical, such as VLBI, shorter Parks–McClellan filters offer a favourable balance. For observations that benefit from a smoother, more stable beam, such as continuum mapping or spectral‑line surveys, longer window‑based designs (e.g., Hamming or Kaiser) are preferable. Cross‑polar leakage is reduced across all designs, improving polarimetric stability.

The framework is inherently adaptable: filter parameters can be adjusted without hardware changes, and the same normalised design can be rescaled to different observing frequencies. Integration into existing calibration pipelines is straightforward, and real‑time implementation on FPGA or GPU back‑ends is feasible, opening the door to adaptive beam shaping.

Several directions for further work emerge naturally. Real‑time implementation would enable dynamic filter adjustment during observations. Machine learning could automate parameter selection from holographic measurements. Extending the method to higher frequencies or different antenna geometries would test its generality. Finally, adaptive filtering that responds to environmental or structural changes could provide continuous beam stabilisation.

In summary, FIR spatial filtering offers a practical, generalisable, and computationally efficient approach to beam optimisation for single‑dish radio telescopes. Its adoption can enhance calibration stability and observational fidelity, complementing traditional mechanical and optical methods. As radio astronomy moves towards more complex and distributed facilities, digital beam‑shaping techniques of this kind will become increasingly valuable for maintaining consistent performance across heterogeneous systems.

\section*{Acknowledgements}

The authors thank the anonymous referees for their constructive and insightful feedback, which has greatly improved the quality of this manuscript. TA--N acknowledges the Ghana Space Science and Technology Institute (GSSTI) of the Ghana Atomic Energy Commission (GAEC) for institutional support and access to its High-Performance Computing (HPC) facilities used in the simulations. 
This research was carried out under the collaborative framework of the Memorandum of Understanding between GAEC, acting through GSSTI, and the Regents of the University of California, Santa Cruz (UCSC). The authors gratefully acknowledge funding and logistical support provided through the GRAO–UCSC Astronomy Development Project, and the technical assistance of colleagues at both GSSTI and the UCSC Department of Astronomy and Astrophysics.

\section*{Data Availability}

The data underlying this article are derived from electromagnetic simulations of the GRAO 32-m telescope conducted using \texttt{GRASP}.  
These data, together with the associated analysis scripts used to generate the beam patterns and filtering results, will be made available to qualified researchers upon reasonable request to the corresponding author.  
No proprietary or confidential information is included.

\section*{Conflict of Interest}

The authors confirm that they have no competing financial interests or personal relationships that could have inappropriately influenced the conduct or reporting of this research. All co-authors have reviewed the final manuscript and unanimously approve its submission for publication. Although this work received financial and logistical support from the GRAO-UCSC Astronomy Development Project, as well as institutional resources from the GSSTI and the GAEC, these funding bodies and institutions were not involved in the study design, data collection, analysis, interpretation of results, or the preparation and revision of this manuscript. The authors bear sole responsibility for the scientific content and the decision to submit this work for publication.



\bibliographystyle{rasti}
\bibliography{example} 








\bsp	
\label{lastpage}
\end{document}